\documentclass[]{article}
\usepackage{multicol}
\usepackage{graphicx}
\usepackage{amssymb}
\usepackage{epsfig}
\usepackage[greek,english]{babel}
\usepackage[iso-8859-7]{inputenc}
\usepackage{feynmp}
\usepackage{subcaption}
\usepackage{amsmath}
\usepackage{float}
\usepackage{verbatim}
\usepackage{wrapfig}
\usepackage{textcomp}
\usepackage{color}
\usepackage{sidecap}
\usepackage{cancel}
\usepackage{amssymb}
\usepackage{kerkis}
\usepackage{youngtab}
\usepackage{stackrel}
\usepackage[toc,page]{appendix}
\usepackage{multirow}
\usepackage{bm}
\usepackage{cite}
\usepackage[absolute]{textpos}
\usepackage[stable]{footmisc}
\newcommand{\by}{by }
\newcommand{\thee}{the }
\newcommand{\as }{as }
\newcommand{\andd }{and }
\newcommand{\newc}{\newcommand}
\newc{\ra}{\rightarrow}
\newc{\lra}{\leftrightarrow}
\newc{\ov}{\overline}
\newc{\pa}{\partial}
\newcommand{\cl}{\mathcal{L}}
\newc{\be}{\begin{equation}}

\newc{\ba}{\begin{eqnarray}}
\newc{\ea}{\end{eqnarray}}
\newc{\n}{\nu}
\newc{\D}{\Delta}

\newc{\la}{\lambda}
\newc{\e}{\epsilon}
\newc{\nn}{\nonumber}

\def\bea{\begin{eqnarray}}
\def\eea{\end{eqnarray}}
\begin{document}
\title{Higher-Dimensional Unification with continuous and fuzzy coset
spaces as extra dimensions\thanks{Based on talks given at: 1)
``Dubna International Advanced School of Theoretical Physics
(DIAS-TH)''. 2) International Conference ``Quantum Field Theory and
Gravity (QFTG'14)'', Tomsk, August 3, 2014. 3) 8th Mathematical
Physics Meeting: ``Summerschool and Conference on modern
Mathematical Physics'', Belgrade, 24-31.04.2014. 4) ``The Joint
Meeting on Quantum Field Theory and Nonlinear Dynamics'', Sinaia,
Romania, 24-28/09/2014 and several seminars by G.Z.}}

\author{D. Gavriil\textsuperscript{1}\,, G. Manolakos\textsuperscript{1}\,,
G. Orfanidis\textsuperscript{1}\,, G.
Zoupanos\textsuperscript{1,2,3}}\date{}
\maketitle
\begin{center}
\emph{E-mails: d.gavriil@yahoo.gr\,,\, gmanol@central.ntua.gr\,,
orfanidi@mail.ntua.gr\,, George.Zoupanos@cern.ch }
\end{center}

\begin{center}
\itshape\textsuperscript{1}Physics Department, National Technical
University,\\ Zografou Campus, GR-15780 Athens, Greece\\
\itshape\textsuperscript{2}Max-Planck-Institut f{\"u}r Physik (Werner-Heisenberg-Institut) \\
F{\"o}hringer Ring 6 80805 M{\"u}nchen \\
\itshape\textsuperscript{3}Arnold-Sommerfeld-Center f{\"u}r
Theoretische Physik Department f{\"u}r Physik,\\
Ludwig-Maximilians-Universit{\"a}t M{\"u}nchen
\end{center}
\vspace{0.1cm} \emph{Keywords}: higher-dimensional theories, coset
space dimensional reduction, fuzzy spheres, orbifold projection

%

\begin{abstract}
We first review the Coset Space Dimensional Reduction (CSDR)
programme and present the best model constructed so far based on the
$\mathcal{N} = 1$, $10$-dimensional $E_8$ gauge theory reduced over
the nearly-K{\"a}hler manifold $SU(3)/U(1)\times U(1)$ with the
additional use of the Wilson flux mechanism. Then we present the
corresponding programme in the case that the extra dimensions are
considered to be fuzzy coset spaces and the best model that has been
constructed in this framework too. In both cases the best model
appears to be the trinification GUT $SU(3)^3$.
\end{abstract}

\section{Introduction}

The unification of fundamental interactions has always been the
theoretical physicists' desideratum. The entire community has set
this aspect in high priority and therefore serious research activity
has led to interesting approaches. Very appealing are the ones that
support the existence of extra dimensions. The extra dimensions'
scenario is encouraged by a very consistent framework, i.e.
superstring theories \cite{green-scwarz-witten}, which, until today,
are considered as best candidates for the unification of all four
fundamental interactions and furthermore are consistently defined
only in ten dimensions. The Heterotic String \cite{gross-harvey} is
regarded as the most promising of all superstring theories, as it
offers a connection to the low-energy physics within the
experimental range, mainly because of the inclusion of a
ten-dimensional $\mathcal{N}=1$ gauge sector. So, compactification
of the ten-dimensional spacetime and dimensional reduction of the
original $E_8\times E_8$ gauge theory, leads to interesting (from a
phenomenological point of view) Grand Unified Theories (GUTs), in
which the Standard Model (SM) could be embedded \cite{gross-harvey}.
Besides the superstring theories, a few years earlier than the
discovery of the Heterotic string, another remarkable framework for
the unification attempt was employed, that is the dimensional
reduction of higher-dimensional gauge theories. Specifically,
Forgacs and Manton with studies on Coset Space Dimensional Reduction
(CSDR) \cite{forgacs-manton, kapetanakis-zoupanos, kubyshin-mourao}
and the Scherk-Schwarz group manifold reduction
\cite{scherk-schwarz}, pioneered this field. In these two
approaches, gauge and Higgs fields in four dimensions are the
surviving components of the gauge fields of an initial gauge theory
defined in higher dimensions, where gauge-Higgs unification is
achieved. Also, in the CSDR scheme, the inclusion of fermions in the
higher-dimensional initial gauge theory, results in Yukawa couplings
in four dimensions. An important step in this unified description is
to demand that the higher-dimensional gauge theory is
$\mathcal{N}=1$ supersymmetric, i.e. the gauge and fermion fields
are members of the same vector supermultiplet, in order to relate
the gauge and fermion fields that have been introduced. The
possibility of obtaining chiral theories in four dimensions
\cite{mantonns,Chapline-Slansky} is regarded as a remarkable
achievement.

In the context of the above framework very welcome suggestions
coming from superstring theory (more specifically from the Heterotic
string \cite {gross-harvey}) are the dimensions of the space-time
and the gauge group of the higher-dimensional supersymmetric theory.
According to the fact that Superstring theory is consistent only in
ten dimensions, the following crucial issues have to be addressed,
(a) distinguish the extra dimensions from the four experimentally
approachable dimensions that we experience, i.e. define a suitable
compactification which is a solution of the theory and (b) determine
the four-dimensional resulting theory. Additionally, using the
appropriate compactification manifolds, one should obtain
$\mathcal{N} = 1$ supersymmetry, having a chance to be led to
realistic GUTs.

In order to preserve an $\mathcal{N}=1$ supersymmetry after
dimensional reduction, Calabi-Yau (CY) spaces serve as suitable
compact internal manifolds \cite {Candelas}. However, the moduli
stabilization problem that arose, led to the study of
compactification with fluxes. Within the context of flux
compactification, the recent developments suggested the use of a
wider class of internal spaces, called manifolds with
SU(3)-structure. The latter class of manifolds admits a
nowhere-vanishing, globally-defined spinor, which is covariantly
constant with respect to a connection with torsion and not with
respect to the Levi-Civita connection as in the CY case. Here we
consider an interesting class of SU(3)-structure manifolds called
nearly-K{\"a}hler manifolds \cite{cardoso-curio, Irges-Zoupanos,
i-z, Butruille}.

The homogeneous nearly-K{\"a}hler manifolds in six dimensions are
the three non-symmetric coset spaces $G_2/SU(3)$,
$Sp(4)/(SU(2)\times U(1))_{non-max}$ and $SU(3)/U(1)\times U(1)$ and
the group manifold $SU(2)\times SU(2)$  \cite{Butruille} (see also
\cite{Irges-Zoupanos, i-z, cardoso-curio}). It is worth noting that
four-dimensional theories resulting from the dimensional reduction
of a ten-dimensional, $\mathcal{N}=1$ supersymmetric gauge theory
over non-symmetric coset spaces, contain automatically supersymmetry
breaking terms \cite{manousselis-zoupanos},
\cite{manousselis-zoupanos2}, contrary to CY spaces. In section 2.3
we present the dimensional reduction of the $\mathcal{N}=1$
supersymmetric $E_8$ gauge theory resulting in the field theory
limit of the Heterotic String over the nearly-K{\"a}hler manifold
$SU(3)/U(1)\times U(1)$. Specifically, an extension of the Minimal
Supersymmetric Standard Model (MSSM) was derived by dimensionally
reducing the $E_8\times E_8$ gauge sector of the Heterotic String
\cite{Irges-Zoupanos, i-z, cardoso-curio}.

Another promising framework which offers a description of physics at
Planck scale is Non-commutative geometry \cite{connes} -
\cite{fuzzy}. Non-commutative geometry was regarded as an
appropriate framework for regularizing quantum field theories, or
even better, building finite ones. Unfortunately, constructing
quantum field theories on Non-commutative spaces is much more
difficult than expected and, moreover, problematic ultraviolet
features have emerged \cite{filk} (see also \cite{grosse-wulkenhaar}
and \cite{grosse-steinacker}). However, this framework is
appropriate to accommodate particle models with Non-commutative
gauge theory \cite{connes-lott} (see also \cite{martin-bondia,
dubois-madore-kerner, madorejz}).

Remarkably, the two frameworks came closer \by realizing that in
M-theory and "open string theory", in \thee presence of a
non-vanishing background antisymmetric field, \thee effective
physics on D-branes can be described \by an Non-commutative gauge
theory \cite{connes-douglas-schwarz, seiberg-witten}. Thus,
Non-commutative field theories emerge as effective description of
string dynamics. Moreover, \thee type IIB superstring theory,
related to the other superstring theories \by certain dualities, in
its conjectured non-perturbative formulation as a matrix model
\cite{ishibasi-kawai}, is a non-commutative theory. Moreover, major
contribution in the framework of non-commutative geometry was made
by Seiberg and Witten \cite{seiberg-witten}. Their study (map
between commutative and non-commutative gauge theories) triggered
notable developments \cite{chaichian, jurco} and, based on them, a
non-commutative version of SM was built \cite{camlet}. Despite the
interest they present as extensions of the SM, those models fail to
troubleshoot the main problem of the SM, that is the presence of
numerous free parameters, due to the ad hoc consideration of Higgs
and Yukawa sectors. Finally, an interesting programme has been
suggested and investigated
\cite{aschieri-madore-manousselis-zoupanos,
aschieri-grammatikopoulos, steinacker-zoupanos,
chatzistavrakidis-steinacker-zoupanos,fuzzy}, which considers the
extra dimensions being non-commutative. This programme overcomes the
ultraviolet/infrared problems of theories defined in non-commutative
spaces in an obvious way. Then it offers the new possibility to
start with an abelian gauge theory defined on the higher-dimensional
space and obtain a non-abelian theory in four dimensions after
dimensional reduction. In addition another spectacular feature of
this programme is that the theories constructed using
non-commutative (fuzzy) manifolds as approximations of the
continuous ones, are renormalizable contrary to all known
higher-dimensional theories. The latter property was examined from
the four-dimensional point of view too, using spontaneous symmetry
breakings, which mimic the results of the dimensional reduction of a
higher-dimensional gauge theory with fuzzy extra dimensions. Finally
chiral realistic theories have been constructed too in this
framework.

In the present review, we are going to deal with higher-dimensional
gauge theories and the four-dimensional models that result after
their dimensional reduction in both frameworks discussed above. Let
us now briefly state the outline of the paper. In the first part, we
introduce the CSDR scheme and reduce dimensionally a $D$-dimensional
Yang-Mills-Dirac theory accordingly. Then we apply the CSDR
programme in the case of a $\mathcal{N}=1$, $E_8$ SYM reduced over
the non-symmetric coset $SU(3)/U(1)\times U(1)$ \cite{Lust-Zoupanos,
Kapetanakis-Zoupanos2, manousselis-zoupanos2}. Then, after recalling
the Wilson flux mechanism for breaking spontaneously a gauge theory,
we demonstrate its successful application in the previous model.

In the second part, we begin with specifying the non-commutative
geometry that is used as internal space throughout the study,
specifically the fuzzy sphere \cite{madore} (as representative of
the fuzzy coset spaces), whose description is given as a matrix
approximation of the ordinary sphere. Then, since our ultimate aim
is to do gauge theory on this non-commutative space, we focus on how
the gauge fields behave on a fuzzy sphere. Afterwards, as a pilot
application, we present a simple dimensional reduction considering a
non-commutative gauge theory on the $M^4\times S^2_N$ space. Since
the four-dimensional model that emerges from the above reduction is
not satisfactory, we proceed by applying a non-trivial dimensional
reduction, the fuzzy version of CSDR. Then, to support further the
claim that the theories built using fuzzy extra dimensions are
renormalizable, we change strategy and instead of reducing to four
dimensions a higher-dimensional theory with fuzzy extra dimensions,
we examine how a four-dimensional gauge theory develops fuzzy
dimensions due to its spontaneous symmetry breaking. In addition, we
address the important problem of chirality in this framework by
applying a $\mathbb{Z}_3$ orbifold projection on an $\mathcal{N} =
4$ SYM theory. Since the vacuum of the orbifolded $\mathcal{N}=4$
SYM theory is vanishing and with vanishing vevs of the scalar
fields, the geometry of the (twisted) fuzzy spheres is offered as
solution with positive vacuum energy, after the inclusion of soft
supersymmetric terms. In this framework, we study the
$\mathcal{N}=4$ SYM four-dimensional theory which is governed by an
$SU(3N)$ gauge symmetry. After the orbifolding, the resulting theory
is $\mathcal{N}=1$ and the gauge group is $SU(N)^3$, since the
orbifold projection induced the breaking of the original gauge
group. The latter does not break in a unique way, but a very
interesting $SU(3)^3$ GUT singles out which is briefly discussed.

\section{The Coset Space Dimensional Reduction.}

In the spirit of the CSDR scheme the obvious way to dimensionally
reduce a gauge theory defined in higher dimensions, is to demand
that the field dependence on the extra coordinates is such that the
Lagrangian is independent of them. While a crude way to fulfill this
requirement is to discard the field dependence on the extra
coordinates, an elegant one is to allow for a non-trivial dependence
on them, in the sense that a symmetry transformation by an element
of the isometry group $S$ of the space formed by the extra
dimensions $B$, corresponds to a gauge transformation. Thus, the
Lagrangian will be independent of the extra coordinates just because
it is gauge invariant. The above requirement is the basis of the
CSDR scheme \cite {forgacs-manton, kapetanakis-zoupanos,
kubyshin-mourao}, which assumes $B$ is a compact coset space, $S/R$.

Consider a Yang-Mills-Dirac Lagrangian, with gauge group $G$,
defined on a $D$-dimensional spacetime $M^D$, with metric $g^{MN}$,
which is compactified to $M^4\times S/R$, with $S/R$ a coset space.
Let the metric have the following form
\begin{equation}
g^{MN}=\left(
         \begin{array}{cc}
           \eta^{\mu \nu} & 0 \\
           0 & -g^{ab} \\
         \end{array}
       \right)\,,
\end{equation}
where $\eta^{\mu \nu}=diag(1,-1,-1,-1)$ and $g^{ab}$ is the coset
space metric. It is important to note that certain constraints on
the fields are imposed by the requirement that transformations of
the fields under the action of the symmetry group of $S/R$ are
compensated by gauge transformations. The solution of these
constraints provides us with the four-dimensional unconstrained
fields, as well as with the gauge invariance that remains in the
theory after dimensional reduction. Along the above framework, a
potential unification of all low energy interactions, gauge, Yukawa
and Higgs is achieved.

The dimensional reduction of higher-dimensional theories results in
effective field theories that might contain also towers of massive
higher harmonic (Kaluza-Klein) excitations. The quantum level
contributions of these excitations alter the behaviour of the
running couplings from logarithmic to power \cite
{Taylor-Veneziano}, resulting in a remarkable change of the
traditional unification picture \cite {Dienes}. Higher-dimensional
theories have also been studied at the quantum level, using the
continuous Wilson renormalization group \cite {Kobayashi-Kubo},
which can be formulated in any number of space-time dimensions with
results in agreement with the treatment involving massive
Kaluza-Klein excitations.

\subsection{Reduction of a $D$-dimensional Yang-Mills-Dirac Lagrangian.}

According to the CSDR framework, the action of the extra dimensions
symmetry group $S$, on the fields defined on $M^4\times S/R$, is
compensated by a gauge transformation. Thus, a gauge invariant
Lagrangian written on this space is independent of the extra
coordinates. Fields defined in this way are called symmetric. The
initial gauge field $A_M (x,y)$ is split into its components $A_\mu
(x,y)$ and $A_\alpha (x,y)$, corresponding to $M^4$ and $S/R$
respectively.

Consider the action of a $D$-dimensional Yang-Mills-Dirac theory
with gauge group $G$ defined on a manifold $M^D$ which, as stated,
will be compactified to $M^4\times  S/R$, $D=4+d$, $d=dimS-dimR$:

 \vspace*{0.3cm}
\begin{align}
A&=\int
d^4xd^dy\sqrt{-g}\left[-\frac{1}{4}Tr(F_{MN}F_{K\Lambda})g^{MK}g^{N\Lambda}+\frac{i}{2}\bar
\psi \Gamma^MD_{M}\psi\right]\,,
\end{align}
\noindent where
\begin{align}
D_{M}=\partial_{M}-\theta_{M}-A_{M},
\end{align}
with
\begin{align}
\theta_{M}=\frac{1}{2}\theta_{MN\Lambda}\Sigma^{N\Lambda},
\end{align}
the spin-connection of $M^{D}$, and
\begin{align}
F_{MN}=\partial_MA_N-\partial_NA_M-[A_M,A_N],
\end{align}
where $M, N$ run over the $D$-dimensional space and $A_M$ and $\psi$
are $D$-dimensional symmetric fields. The fermion fields can be
accommodated in any representation $F$ of $G$, unless a further
symmetry, such as supersymmetry, is required. Let $\xi^{\alpha}_{A},
(A=1,...,dimS$ and $\alpha =dimR+1,...,dimS$ the curved index$)$ be
the Killing vectors which generate the symmetries of $S/R$ and $W_A$
the compensating gauge transformation associated with  $\xi_A$. The
following constraint equations for scalar $\phi$, vector
$A_{\alpha}$ and spinor $\psi$ fields on $S/R$, are expressing the
requirement that transformations of the fields under the action of
$S/R$  are compensated by gauge transformations
\begin{gather}
\delta_A\phi=\xi^{\alpha}_{A}\partial_\alpha\phi=D(W_A)\phi,\label{ena} \\
\delta_AA_{\alpha}=\xi^{\beta}_{A}\partial_{\beta}A_{\alpha}+\partial_{\alpha}\xi^{\beta}_{A}A_{\beta}=\partial_{\alpha}W_A-[W_A,A_{\alpha}], \label{duo}\\
\delta_A\psi=\xi^{\alpha}_{A}\partial_\alpha\psi-\frac{1}{2}G_{Abc}\Sigma^{bc}\psi=D(W_A)\psi\,,
\label{tria}
\end{gather}
where $W_A$ depend only on internal coordinates $y$ and $D(W_A)$
represents a gauge transformation in the appropriate representation
of the fields.

We obtain \cite {forgacs-manton, kapetanakis-zoupanos} the
four-dimensional unconstrained fields, as well as the gauge
invariance that remains in the theory after dimensional reduction,
by the detailed analysis of the constraints
\eqref{ena}-\eqref{tria}. The components $A_\mu (x,y)$ of the
initial gauge field $A_M (x,y)$ become, after dimensional reduction,
the four dimensional gauge fields and they are independent of $y$.
Moreover, one can find that they commute with the generators of
$R_G$, subgroup of $G$. In other words, the four-dimensional gauge
group $H$ is the centralizer of $R$ in $G$, $H=C_G(R_G)$. The
$A_{\alpha}(x,y)$ components of $A_M(x,y)$ denoted by
$\phi_{\alpha}(x,y)$ from now on, become scalars in four dimensions
and they transform under $R$ as a vector $\upsilon$, i.e.
\begin{gather}
S\supset R \\
adjS=adjR+\sum(s_i)\,.
\end{gather}
In addition, $\phi_{\alpha}(x,y)$ act as an interwining operator
connecting induced representations of $R$ acting on $G$ and $S/R$.
This implies that in order to find the representation of the gauge
group $H$ under which the $\phi$'s transform in four dimensions, we
have to decompose $G$ according to the embedding:
\begin{gather}
G\supset R_G\times H \label{tria_ena}\\
adjG=(adjR,1)+(1,adjH)+\sum(r_i,h_i)\,. \label{tria_duo}
\end{gather}
Then, for every pair $r_i,s_i$, where $r_i$ and $s_i$ are identical
irreducible representations of $R$, there survives a Higgs multiplet
transforming under the representation $h_i$ of $H$. All other scalar
fields vanish.

Regarding the fermion fields \cite{kapetanakis-zoupanos,
Wetterich-Palla, mantonns,Chapline-Slansky} we proceed along similar
lines as in the case of scalars. It turns out that the spinor fields
act as interwining operator connecting induced representation of $R$
in $SO(d)$ and $G$. Similar to the scalar cases, in order to specify
the representation of $H$ under which the four-dimensional fermions
transform, we have to consider the initial gauge group and decompose
its representation $F$, in which the fermions are assigned in higher
dimensions, under $R_G\times H$, i.e.
\begin{align}
\begin{split}
F=\sum (t_i,h_i)
\end{split}
\end{align}
and the spinor of $SO(d)$ under $R$
\begin{align}
\begin{split}
\sigma_d=\sum \sigma _j\,.
\end{split}
\end{align}
Then for each pair $t_i$ and $s_i$, where $t_i$ and $s_i$ are
identical irreducible representations of $R$, there is  an $h_i$
multiplet of spinor fields in the four-dimensional theory. Regarding
the existence of chiral fermions in the effective theory, we notice
that if we start with Dirac fermions in higher dimensions it is
impossible to obtain chiral fermions in four dimensions. We have to
impose further requirements in order to achieve chiral fermions in
the resulting theory. Imposing the Weyl condition in $D$ dimensions,
we obtain two sets of Weyl fermions with the same quantum numbers
under $H$. Although we already have a chiral theory, we can go
further and try to impose the Majorana condition in order to
eliminate the doubling of the fermionic spectrum. Majorana and Weyl
conditions are compatible in $D=4n+2$, which is the case of our
interest.

An important requirement is that the resulting four-dimensional
theories should be anomaly free. Along that direction, Witten \cite
{Witten} starting with an anomaly free theory in high dimensions,
has given the condition to be fulfilled in order to obtain anomaly
free theories in four dimensions after the dimensional reduction.
The condition restricts the allowed embeddings of $R$ into $G$ by
relating them with the embedding of $R$ into $SO(6)$, the tangent
space of the six-dimensional cosets we consider \cite
{kapetanakis-zoupanos,Piltch}. According to ref. \cite {Piltch}, the
anomaly cancelation condition is automatically satisfied for the
choice of embedding
\begin{align}
\begin{split}
E_8\supset SO(6)\supset R,
\end{split}
\end{align}
which we adopt here.

\subsection{The Four-Dimensional Theory}

In order to obtain the four-dimensional effective action, we take
into account all the constraints and we integrate out the extra
coordinates. We assume that the metric is block diagonal and we have
the following Lagrangian: \vspace*{0.3cm}
\begin{align}
A&=C\int d^4x[-\frac{1}{4}F_{\mu \nu}^tF^{t\mu
\nu}+\frac{1}{2}(D_\mu \phi _\alpha)^t(D^\mu \phi
^\alpha)^t+V(\phi)+\frac{i}{2}\bar \psi
\Gamma^{\mu}D_{\mu}\psi-\frac{i}{2}\bar \psi
\Gamma^{a}D_{a}\psi]\,,\label{tessera}
\end{align}
where $D_\mu = \partial_{\mu}-A_\mu$ and $D_a =
\partial_{a}-\theta_a-\phi_a$, with $\theta_a =
\frac{1}{2}\theta_{abc}\Sigma^{bc}$ the connection of the space,
while $C$ is the volume of the space. The potential $V(\phi)$ is
given by the following expression
\begin{align}
V(\phi)=-\frac{1}{4}g^{ac}g^{bd}Tr(f^C_{ab} \phi_C -[\phi_a,
\phi_b])(f^D_{cd} \phi_D - [\phi_c, \phi_d]),
\end{align}
where, $A=1,...,dimS$ and $f$'s are the structure constants
appearing in the commutators of the Lie algebra of $S$. The scalar
fields $\phi_a$ appearing in $V(\phi)$ must satisfy the following
equation coming from  \eqref{ena}-\eqref{tria}
\begin{align}
f^D_{ai}\phi_D-[\phi_a, \phi_i]=0\,, \label{tessera_ena}
\end{align}
where the $\phi_i$ generate $R_G$. As a consequence, some of the
$\phi_a$'s might not survive the dimensional reduction procedure,
while the rest can be identified with the genuine Higgs fields.
Expressing $V(\phi)$ in terms of the unconstrained independent Higgs
fields, it remains a quartic polynomial, which is invariant under
the four-dimensional gauge group $H$. The unbroken final group is
determined by the minimization of the potential
\cite{Harnad,Chapline-Manton,Farakos-Koutsoumbas}. Although this is
generally a difficult task, there is a special case in which things
turn out to be quite simple. This is the case when $S$ has an
isomorphic image $S_G$ in $G$ which contains $R_G$ in a consistent
way. It is possible then to allow the $\phi_a$ to become generators
of $S_G$. That is $\overline{\phi}_a=<\phi^i>Q_{ai}=Q_a$ with
$<\phi^i>Q_{ai}$ suitable combinations of $G$ generators of $S_G$
and $a$ is also a coset space index. Therefore, because of the
commutation relations of $S$, we find
\begin{align*}
\overline{F}_{ab} &= f_{ab}^iQ_i+f_{ab}^c\overline{\phi}_c-[\overline{\phi}_a,\overline{\phi}_b]\\
&= f_{ab}^iQ_i+f_{ab}^cQ_c-[Q_a,Q_b]=0.
\end{align*}
Thus, we have proven that $V(\phi=\overline{\phi})=0$ and because
$V$ is positive definite, is also the minimum of the potential.
These non-zero vacuum expectation values of the Higgs fields, break
further the four-dimensional gauge group $H$ to the centralizer $K$
of the image of $S$ in $G$, i.e. $K=C_G(S)$
\cite{kapetanakis-zoupanos,Harnad,Chapline-Manton,Farakos-Koutsoumbas}.
The above statement can be checked by the gauge transformation
\begin{align*}
\phi_a\to h\phi_ah^{-1},h\in H,
\end{align*}
noting that the vev of the Higgs fields is gauge invariant for the
set of $h$'s that commute with $S$.

Regarding the fermion part of the Lagrangian, the first term is just
the kinetic term of fermions, while the second is the Yukawa term
\cite{Kapetanakis-Zoupanos2}. The representation in which the
fermions are assigned under the gauge group can be real, if $\psi$
is a Majorana-Weyl spinor in ten dimensions. The last term in
\eqref{tessera} can be written as
\begin{align}
L_Y = -\frac{i}{2}\bar \psi \Gamma^{a}(\partial_a
-\frac{1}{2}f_{ibc}e_\Gamma^i e_a^\Gamma \Sigma^{bc} -
\frac{1}{2}G_{abc}\Sigma^{bc}-\phi_a)\psi=\frac{i}{2} \bar \psi
\Gamma^a \nabla_a \psi +\bar \psi V\psi\,,\label{pente}
\end{align}
where
\begin{align}
\nabla_a &=-\partial_a +\frac{1}{2}f_{ibc}e_\Gamma^i e_a^\Gamma \Sigma^{bc}+\phi_a, \label{exi} \\
V &=\frac{i}{4}\Gamma^aG_{abc}\Sigma^{bc},
\end{align}
using the full connection with torsion \cite{Muller-Hoissen}. The
fermion fields are independent of the coset space coordinates i.e.
$\partial_a\psi=0$ since they are symmetric fields, satisfying the
constraint equation \eqref{tria}. Furthermore we can consider the
Lagrangian at the point $y=0$, due to its invariance under
$S$-tranformations, and $e^i_{\Gamma}=0$ at that point. Therefore
\eqref{exi} becomes just $\nabla_a=\phi_a$ and the term
$\frac{i}{2} \bar \psi \Gamma^a \nabla_a \psi$ in  \eqref{pente} is
exactly the Yukawa term.

The examination of the last term appearing in \eqref{pente} reveals
that the operator $V$ anticommutes with the six-dimensional helicity
operator \cite{kapetanakis-zoupanos}. Moreover one can show that $V$
commutes with the $T_i=-\frac{1}{2}f_{ibc}\Sigma^{bc}$ ($T_i$ close
the $R$-subalgebra of $SO(6)$). Exploiting Schur's lemma, we can
draw the conclusion, that the non-vanishing elements of $V$ are only
those which appear in the decomposition of both $SO(6)$ irreps $4$
and $\bar4$ , e.g. the singlets. Since this term is of pure
geometric nature, we reach the conclusion that the singlets in $4$
and $\bar{4}$ will acquire large geometrical masses, a fact that has
serious phenomenological implications. Although the above conclusion
in the higher-dimensional supersymmetric theory framework means that
the gauginos obtained in four dimensions after the dimensional
reduction receive masses comparable to the compactification scale,
this result changes in presence of torsion.

\subsection{Dimensional Reduction of $E_8$ over $SU(3)/U(1)\times
U(1)$}

Let us next present a few results concerning the dimensional
reduction of the $\mathcal{N}=1$, $E_8$ SYM over the non-symmetric
coset $SU(3)/U(1)\times U(1)$ \cite{Lust-Zoupanos,
Kapetanakis-Zoupanos2}. According to \eqref{tria_ena}, in order to
determine the four-dimensional gauge group we examine the embedding
of $R=U(1)\times U(1)$ in $E_8$ by the decomposition
\begin{align}
E_8\supset E_6\times SU(3)\supset E_6\times U(1)_A \times U(1)_B
\label{efta}
\end{align}
Then the four-dimensional gauge group after dimensional reduction of
$E_8$ under $SU(3)/U(1)\times U(1)$ is given by
\begin{align}
H=C_{E_8}(U(1)_A\times U(1)_B)=E_6\times U(1)_A\times U(1)_B\,.
\end{align}
The explicit decomposition of the adjoint representation of $E_8$,
248 under $U(1)_A\times U(1)_B$ provides us with the surviving
scalars and fermions in four dimensions. Applying \eqref{tria_duo}
to the case of our interest, we result in the following
decomposition
\begin{gather}
248=1_{(0,0)}+1_{(0,0)}+1_{(3,\frac{1}{2})}+1_{(-3,\frac{1}{2})}+1_{(0,-1)}+1_{(0,1)}+1_{(-3,-\frac{1}{2})}+1_{(3,-\frac{1}{2})}+\nonumber\\
78_{(0,0)}+27_{(3,\frac{1}{2})}+27_{(-3,\frac{1}{2})}+27_{(0,-1)}+
\overline {27}_{(-3,-\frac{1}{2})}+
\overline{27}_{(3,-\frac{1}{2})}+\overline{27}_{(0,1)}\,.
\label{okto}
\end{gather}
The $R=U(1)\times U(1)$ decompositions of the vector and spinor
representations of $SO(6)$ are
\begin{align*}
(3,\frac{1}{2})+(-3,\frac{1}{2})+(0,-1)+(-3,-\frac{1}{2})+(3,-\frac{1}{2})+(0,1)
\end{align*}
and
\begin{align*}
(0,0)+(3,\frac{1}{2})+(-3,\frac{1}{2})+(0,-1)\,,
\end{align*}
respectively. Applying the CSDR rules, we find that the surviving
fields in four dimensions are three  $\mathcal{N}=1$ vector
supermultiplets containing the gauge fields of $E_6\times
U(1)_A\times U(1)_B$. Regarding the matter content of the effective
theory, we result in six chiral multiplets, where three of them are
$E_6$ singlets carrying $U(1)_A\times U(1)_B$ charges, while the
remaining are the $A^i, B^i$ and  $C^i$ chiral multiplets, with $i$
an $E_6$, 27 index.

Next we examine the decomposition of the adjoint of the specific
$S=SU(3)$ under $R=U(1)\times U(1)$  i.e. $SU(3)\supset U(1)\times
U(1)$:
\begin{equation}
8=(0,0)+(0,0)+(3,\frac{1}{2})+(-3,\frac{1}{2})+(0,-1)+(-3,-\frac{1}{2})+(3,-\frac{1}{2})+(0,1)\,.
\end{equation}
The above decomposition suggests the introduction of the following
generators for $SU(3)$:
\begin{align}
Q_{SU(3)}=\big\{Q_0,Q_0^{'} ,Q_1,Q_2,Q_3,Q^1,Q^2,Q^3 \big\}.
\label{okto_duo}
\end{align}
The non trivial commutator relations of $SU(3)$ generators,
\eqref{okto_duo}, are given in \cite{Manousselis}. According to this
decomposition, the following notation of the scalar fields is
suggested
\begin{align}
(\phi_I, I=1,...,8)\longrightarrow
(\phi_0,\phi_0^{'},\phi_1,\phi^1,\phi_2,\phi^2,\phi_3,\phi^3).
\label{ennia}
\end{align}
In terms of the redefined fields \eqref{ennia}, the potential of any
theory reduced over $SU(3)/U(1)\times U(1)$ is given by
\begin{multline}
V(\phi)=(3\Lambda^2+\Lambda^{'2})(\frac{1}{R_1^4}+\frac{1}{R_2^4})+\frac{4\Lambda^{'2}}{R_3^2}\\
+\frac{2}{R_2^2R_3^2}Tr(\phi_1\phi^1)+\frac{2}{R_1^2R_3^2}Tr(\phi_2\phi^2)+\frac{2}{R_1^2R_2^2}Tr(\phi_3\phi^3)\\
+\frac{\sqrt{3}\Lambda}{R_1^4}Tr(Q_0[\phi_1,\phi^1])-\frac{\sqrt{3}\Lambda}{R_2^4}Tr(Q_0[\phi_2,\phi^2])-\frac{\sqrt{3}\Lambda}{R_3^4}Tr(Q_0[\phi_3,\phi^3]) \\
+\frac{\Lambda^{'}}{R_1^4}Tr(Q_0^{'}[\phi_1,\phi^1])+\frac{\Lambda^{'}}{R_2^4}Tr(Q_0^{'}[\phi_2,\phi^2])-\frac{2\Lambda^{'}}{R_3^4}Tr(Q_0^{'}[\phi_3,\phi^3]) \\
+\Big[\frac{2\sqrt{2}}{R_1^2R_2^2}Tr(\phi_3[\phi_1,\phi_2])+\frac{2\sqrt{2}}{R_1^2R_3^3}Tr(\phi_2[\phi_3,\phi_1])+\frac{2\sqrt{2}}{R_2^2R_3^2}Tr(\phi_1[\phi_2,\phi_3])+h.c\Big] \\
+\frac{1}{2}Tr\Big(\frac{1}{R_1^2}([\phi_1,\phi^1])+\frac{1}{R_2^2}([\phi_2,\phi^2])+\frac{1}{R_3^2}([\phi_3,\phi^3])\Big)^2\\-\frac{1}{R_1^2R_2^2}Tr([\phi_1,\phi_2][\phi^1,\phi^2])-\frac{1}{R_1^2R_3^2}Tr([\phi_1,\phi_3][\phi^1,\phi^3])-\frac{1}{R_2^2R_3^2}Tr([\phi_2,\phi_3][\phi^2,\phi^3]),
\label{ennia_duo}
\end{multline}
where $R_1$, $R_2$, $R_3$ are the coset space radii\footnote{To
bring the potential into this form we have used (A.22) of
ref.\cite{kapetanakis-zoupanos} and relations (7),(8) of
ref.\cite{Witten2}.}. The real metric \footnote{The complex metric
that was used is
$g^{1\bar{1}}=\frac{1}{R_1^2},g^{2\bar{2}}=\frac{1}{R_2^2},g^{2\bar{3}}=\frac{1}{R_3^2}$.}
of the coset is expressed in terms of the radii by
\begin{align}
g_{ab}=diag(R_1^2,R_1^2,R_2^2,R_2^2,R_3^2,R_3^2).
\end{align}
Under the decomposition \eqref{okto} the generators of $E_8$ can be
grouped as
\begin{align}
Q_{E_8}=\big\{Q_0,Q_0^{'}
,Q_1,Q_2,Q_3,Q^1,Q^2,Q^3,Q^{\alpha},Q_{1i},Q_{2i},Q_{3i},
Q^{1i},Q^{2i},Q^{3i} \big\},
\end{align}
where $\alpha =1,...,78$ and $i=1,...,27$. The redefined fields in
\eqref{tessera_ena} are subject to the constraints
\begin{align}
[\phi_1,\phi_0] &= \sqrt{3}\phi_1, & [\phi_3,\phi_0]&=0, & [\phi_1,\phi_0^{'}]&=\phi_1, \nonumber\\
[\phi_2,\phi_0] &= -\sqrt{3}\phi_2, & [\phi_2,\phi_0^{'}]&=\phi_2, &
[\phi_3,\phi_0^{'}]&=-2\phi_3\,.
\end{align}
The solutions of the constraints in terms of the genuine Higgs
fields and the $E_8$ generators corresponding to the embedding of
$R=U(1)\times U(1)$ in the $E_8$ are, $\phi_0=\Lambda Q_0$ and
$\phi_0^{'}=\Lambda^{'}  Q_0^{'}$, with $ \Lambda =\Lambda^{'}
=\frac{1}{\sqrt{10}}$ and
\begin{align*}
\phi_1 &= R_1 \alpha^i Q_{1i}+R_1 \alpha Q_1, & \phi_2 &= R_2
\beta^i Q_{2i}+R_2\beta Q_2, & \phi_3 &= R_3\gamma^iQ_{3i}+R_3\gamma
Q_3,
\end{align*}
where the unconstrained fields transform under $E_6\times
U(1)_A\times U(1)_B$ as
\begin{align*}
\alpha_i \sim 27_{(3,\frac{1}{2})}, \quad  \beta_i \sim
27_{(-3,\frac{1}{2})}, \quad  \gamma_i \sim 27_{(0,-1)}, \quad
\alpha \sim 1_{(3,\frac{1}{2})}, \quad  \beta \sim
1_{(-3,\frac{1}{2})}, \quad  \gamma \sim 1_{(0,-1)}.
\end{align*}
The scalar potential \eqref{ennia_duo}  becomes
\begin{multline}
V(\alpha^i,\alpha,\beta^i,\beta,\gamma^i,\gamma)= const.+\bigg(\frac{4R_1^2}{R_2^2 R_3^2}-\frac{8}{R_1^2}\bigg)\alpha^i \alpha_i + \bigg(\frac{4R_1^2}{R_2^2 R_3^2}- \frac{8}{R_1^2}\bigg)\bar{\alpha} \alpha \\
+\bigg(\frac{4R_2^2}{R_1^2 R_3^2}-\frac{8}{R_2^2}\bigg)\beta^i \beta_i +\bigg(\frac{4R_2^2}{R_1^2 R_3^2}-\frac{8}{R_2^2}\bigg)\bar{\beta} \beta \\
+\bigg(\frac{4R_3^2}{R_1^2 R_2^2}-\frac{8}{R_3^2}\bigg)\gamma^i \gamma_i +\bigg(\frac{4R_3^2}{R_1^2 R_2^2}-\frac{8}{R_3^2}\bigg)\bar{\gamma} \gamma\\
+\bigg[\sqrt{2}80 \bigg(\frac{R_1}{R_2 R_3}+\frac{R_2}{R_1 R_3}+\frac{R_3}{R_2 R_1}\bigg)d_{ijk}\alpha^i \beta^j \gamma^k \\
+\sqrt{2}80\bigg(\frac{R_1}{R_2 R_3}+\frac{R_2}{R_1 R_3}+\frac{R_3}{R_2 R_1}\bigg)\alpha \beta \gamma +h.c \bigg] \\
+ \frac{1}{6}\bigg( \alpha^i(G^\alpha)_i^j\alpha_j+\beta^i(G^\alpha)_i^j\beta_j+\gamma^i(G^\alpha)_i^j\gamma_j\bigg)^2 \\
+\frac{10}{6}\bigg( \alpha^i(3\delta_i^j)\alpha_j+\bar{\alpha}(3)\alpha+\beta^i(-3\delta_i^j)\beta_j+\bar{\beta}(-3)\beta \bigg)^2 \\
+\frac{40}{6}\bigg( \alpha^i(\frac{1}{2}\delta_i^j)\alpha_j+\bar{\alpha}(\frac{1}{2})\alpha+\beta^i(\frac{1}{2}\delta_i^j)\beta_j+\bar{\beta}(\frac{1}{2})\beta+\gamma^i(-1\delta_i^j)\gamma^j+\bar{\gamma}(-1)\gamma \bigg)^2 \\
+40\alpha^i \beta^j d_{ijk}d^{klm} \alpha_l \beta_m+40\beta^i
\gamma^j d_{ijk}d^{klm} \beta_l
\gamma_m+40 \alpha^i \gamma^jd_{ijk} d^{klm} \alpha_l \gamma_m \\
+40(\bar{\alpha}\bar{\beta})(\alpha\beta)+40(\bar{\beta}\bar{\gamma})(\beta\gamma)+40(\bar{\gamma}\bar{\alpha})(\gamma
\alpha)\label{deka}
\end{multline}
and it is positive definite. From the above potential we read the
$F-,D-$ and scalar soft terms. The $F$ terms are obtained from the
superpotential
\begin{align}
\mathcal{W}(A^i,B^j,C^k,A,B,C)=\sqrt{40}d_{ijk}A^iB^jC^k+\sqrt{40}ABC\,.
\end{align}
The $D$-terms have the following structure
\begin{align}
\frac{1}{2}D^{\alpha}D^{\alpha}+\frac{1}{2}D_1D_1+\frac{1}{2}D_2D_2\,,
\end{align}
where
\begin{align*}
D^{\alpha}=\frac{1}{\sqrt{3}}\Big(\alpha^i(G^{\alpha})_i^j\alpha_j+\beta^i(G^{\alpha})_i^j\beta_j+\gamma^i(G^{\alpha})_i^j\gamma_j\Big),\\
D_1=\sqrt{\frac{10}{3}}\Big(\alpha^i(3\delta_i^j)\alpha_j+\bar{\alpha}(3)\alpha+\beta^i(-3\delta_i^j)\beta_j+\bar{\beta}(-3)\beta\Big)
\end{align*}
and
\begin{align*}
D_2=\sqrt{\frac{40}{3}}\Big(\alpha^i(\frac{1}{2}\delta_i^j)\alpha_j+\bar{\alpha}(\frac{1}{2})\alpha+\beta^i(\frac{1}{2}\delta_i^j)\beta_j+\bar{\beta}(\frac{1}{2})\beta+\gamma^i(-1\delta_i^j)\gamma_j+\bar{\gamma}(-1)\gamma\Big)\,.
\end{align*}
The rest of the terms in the potential could be interpreted as soft
scalar masses and trilinear soft terms. Finally, the gaugino obtains
a mass of the order $\mathcal{O}(R^{-1})$, and receives contribution
from the torsion contrary to the rest soft supersymmetric breaking
terms.

Concluding the present subsection, we would like to note the fact
that, starting with an $\mathcal{N} = 1$ supersymmetric theory in
ten dimensions, the CSDR leads to an $\mathcal{N} = 1$ , $E_6$ GUT
with $U(1)_A, U(1)_B$ resulting as global symmetries.

\section{Breaking by Wilson Flux mechanism}
According to the findings of the previous section, the $E_6\times
U(1) \times U(1)$ group is the surviving gauge group of the
initial's $E_8$ group dimensional reduction. In the context of
symmetry reduction, we notice that the $E_6$ gauge group can not
break entirely with the presence of the 27s Higgs representations
only. In order to further reduce the gauge symmetry we employ the
Wilson flux breaking mechanism  \cite{Kozimirov,Zoupanos,Hosotani}.
Let us briefly recall the Wilson flux mechanism for breaking
spontaneously a gauge theory, and present the way that this
mechanism is applied to our model.
\subsection{The Wilson Flux mechanism}
According to the quantum mechanical phenomenon Aharonov-Bohm effect,
the wave function of electrons in the double slit experiment,
experiences  a measurable phase shift $\varphi$, as a result of
their motion in a multiply connected region. The phase shift
$\varphi$, is given by
\begin{equation}
\varphi=\oint_{\gamma} A_i dx^i, \label{enteka}
\end{equation}
where $\gamma$ is a closed path around the solenoid of the
experiment, and the solenoid is considered to be a hole in the
manifold.

In the case of our interest, instead of considering the simply
connected manifold $B_0$, where $B_0$ is the coset $S/R$, we
consider the multiply connected manifold $B=B_0/F^{S/R}$ with
$F^{S/R}$ a freely acting discrete symmetry of $B_0$. The manifold
$B$ is multiply connected due to the presence of the symmetry
$F^{S/R}$. For each element $g\in F^{S/R}$, we pick up an element
$U_g$ in $H$, which can be represented as the Wilson loop (the
generalization of  \eqref{enteka} to non abelian theories)
\begin{equation}
U_g={\mathcal{P}}exp\left(-i\oint_{\gamma_g} T^a A^a_M dx^M \right),
\end{equation}
where $A^a_M$ are vector fields with group generators $T^a$,
$\gamma_g$ is a contour representing the abstract element $g$ of
$F^{S/R}$, and $\mathcal{P}$ denotes the path ordering.  If the
manifold is simply connected, then the vanishing of the field
strength ensures that we can set the gauge field to zero by a gauge
transformation. If $\gamma$ is chosen not to be contractible to a
point, then the Wilson Loop $U[\gamma]\neq 1$ and is gauge
covariant.  In this case, although the vacuum field strength
vanishes everywhere, $U_g$ cannot be set to one and the gauge field
cannot be set to zero. Therefore, a homomorphism of $F^{S/R}$ into
$H$ is induced with image $T^H$, which is the subgroup of $H$
generated by {$U_g$}. Moreover, a field $f(x)$ on $B_0$ is obviously
equivalent to another field on $B_0$ which obeys $f(g(x))=f(x)$ for
every $g\in F^{S/R}$. However the presence of the gauge group $H$
generalizes this statement to
\begin{align}
f(g(x))=U_gf(x).\label{dwdeka}
\end{align}
Concerning the gauge symmetry that is preserved by the vacuum, we
consider the following. The vacuum has $A_\mu^a=0$ and we represent
a gauge transformation by a space-dependent matrix $V(x)$ of $H$. In
order to keep  $A_\mu^a=0$ and leave the vacuum invariant, $V(x)$
must be a constant. Moreover, $f\to Vf$ is consistent with
\eqref{dwdeka} only if $[V,U_g]=0$ for all $g\in F^{S/R}$.
Therefore, the unbroken subgroup of $H$ is the centralizer of $T^H$
in $H$. Concerning the matter fields,  in order to satisfy
\eqref{dwdeka} and therefore survive in the theory, they have to be
invariant under the diagonal sum
\begin{align*}
F^{S/R}\oplus T^H.
\end{align*}
The discrete symmetries $F^{S/R}$, which act freely on coset spaces
$B_0=S/R$ are the center of $S$, $Z(S)$ and $W=W_S/W_R$ , where
$W_S$ and $W_R$ are the Weyl groups of $S$ and $R$, respectively. In
the case of our interest, where  $B_0=SU(3)/U(1)\times U(1)$, we
have
\begin{equation}
F^{S/R}=\mathbb{Z}_3  \subseteq W.
\end{equation}

\subsection{Application to our model}
According to the above, the projected theory in the presence of the
Wilson loop is derived by keeping the invariant under the combined
action of the discrete group $\mathbb{Z}_3$, on the geometry and on
the gauge indices, fields. The $\mathbb{Z}_3$ acts non-trivially on
the various fields of the theory and its action on the gauge indices
is implemented by the matrix \cite{fuzzy}
$\gamma_3=diag\{\mathbf{1}_3,\omega \mathbf{1}_3, \omega^2
\mathbf{1}_3\}$, where $\omega=e^{i\frac{2\pi}{3}}$ acts on the
$E_6$ gauge fields and a non trivial phase acts on the matter
fields. Specifically the gauge fields of $E_6$ that survive the
projection are those that satisfy
\begin{equation}
[A_M,\gamma_3]=0\;\;\Rightarrow A_M=\gamma_3 A_M \gamma_3^{-1}.
\end{equation}
The surviving gauge group after the $\mathbb{Z}_3$ projection is
$$SU(3)_c\times SU(3)_L \times SU(3)_R.$$ The surviving matter
superfields are those that satisfy
\begin{equation}
\vec{a}=\omega\gamma_3\vec{a},\;\;\vec{b}=\omega^2
\gamma_3\vec{b},\;\;\vec{c}=\omega^3 \gamma_3\vec{c}\,,
\end{equation}
where $\vec a,\vec b, \vec c$ are   the $\mathbf {27}$
representation matter superfields.  According to the decomposition
of the  $\mathbf {27}$ representation of $E_6$ under $SU(3)_c\times
SU(3)_L \times SU(3)_R$,  the above superfields are consisted of
three sets of fields,  expressed by the following representations
\begin{equation}
(3,\bar 3,1),\;(1,3,\bar 3)\;,(\bar3,1,3).
\end{equation}
Moreover, the projection on the singlets of $E_6$, is
\begin{equation}
a=\omega a,\;\;b=\omega^2 {b},\;\;{c}=\omega^3 {c}\;,
\end{equation}
and the scalar matter fields are in the bi-fundamental
representations
\begin{align}
a_3\sim H_1\sim (\bar{3},1,3)_{(3,\frac{1}{2})}, \;\; b_2\sim
H_2\sim (3,\bar{3},1)_{(-3,\frac{1}{2})}, \;\; c_1\sim H_3\sim
(1,\bar{3},3)_{(0,-1)}.
\end{align}
One can obtain three fermion generations by introducing non-trivial
monopole charges in the $U(1)$'s in $R$. We denote the resulting
three copies of the bi-fundamental fields as (the index $l=1,2,3 $
specifies the flavours)
\begin{align*}
3\cdot H_1 &\rightarrow H_1^{(l)}\sim 3 \cdot (\bar{3},1,3)_{(3,\frac{1}{2})}, \\
3\cdot H_2 &\rightarrow H_2^{(l)}\sim 3 \cdot (3,\bar{3},1)_{(-3,\frac{1}{2})}, \\
3\cdot H_3 &\rightarrow H_3^{(l)}\sim 3 \cdot
(1,3,\bar{3})_{(0,-1)}.
\end{align*}
Similarly, we denote the three copies of the scalars as
\begin{align}
3\cdot \gamma_{(0,-1)}\rightarrow \theta_{(0,-1)}^{(l)}.
\end{align}
The scalar potential can be rewritten as \cite{Irges:2011de}
\begin{align}
V_{sc}=3(3\Lambda^2+\Lambda^{'2})\Big(\frac{1}{R_1^4}+\frac{1}{R_2^4}\Big)+\frac{3\cdot
4\Lambda^{'2}}{R_3^4}+\underset{l=1,2,3}{\sum}V^{(l)},
\end{align}
where
\begin{align}
V^{(l)}=V_{susy}+V_{soft}\,,
\end{align}
with $V_{susy}=V_D+V_F$. We can drop the flavour superscript $(l)$
from most of the fields, at least until we give vevs to the Higgses
(which in general can be different for each $l$), since there are
three identical contributions to the potential. Then, the explicit
forms of the $D$ and $F$ terms are
\begin{align}
V_D =\frac{1}{2}\underset{A}{\sum}D^AD^A+\frac{1}{2}D_1D_1+\frac{1}{2}D_2D_2, \\
V_F =\underset{i=1,2,3}{\sum}|F_{H_i}|^2+|F_{\theta}|^2, \;\;
F_{H_i}=\frac{\partial\mathcal{W}}{\partial H_i} ,\;\;
F_{\theta}=\frac{\partial\mathcal{W}}{\partial \theta}.
\end{align}
The $F$-terms derive from
\begin{align}
\mathcal{W}=\sqrt{40}d_{abc}H_1^aH_2^bH_3^c
\end{align}
and the D-terms are
\begin{align*}
D^A &=\frac{1}{\sqrt{3}}\big<H_i|G^A|H_i\big>, \\
D_1 &=3\sqrt{\frac{10}{3}}(\big< H_1|H_1\big>-\big<H_2|H_2\big>), \\
D_2 &=\sqrt{\frac{10}{3}}(\big<
H_1|H_1\big>+\big<H_2|H_2\big>-2\big<H_3|H_3\big>-2|\theta|^2) ,
\end{align*}
where
\begin{align*}
\big<H_i|G^A|H_i\big> &= \underset{i=1,2,3}{\sum}H_i^a(G_A)_a^bH_{ib}, \\
\big<H_i|H_i\big> &=  \underset{i=1,2,3}{\sum}H_i^a\delta_a^bH_{ib}.
\end{align*}
Finally, the soft breaking terms are
\begin{multline}
V_{soft}=\big(\frac{4R_1^2}{R_2^2R_3^2}-\frac{8}{R_1^2}\big)\big<H_1|H_1\big>+\big(\frac{4R_2^2}{R_1^2R_3^2}-\frac{8}{R_2^2}\big)\big<H_2|H_2\big> \\
+\big(\frac{4R_3^2}{R_1^2R_2^2}\Big)-\frac{8}{R_3^2}(\big<H_3|H_3\big>+|\theta|^2) \\
+80\sqrt{2}\big(\frac{R_1}{R_2R_3}+\frac{R_2}{R_1R_3}+\frac{R_3}{R_1R_2}\big)(d_{abc}H_1^aH_2^bH_3^c+h.c).
\end{multline}
The $(G^A)_a^b$ are the structure constants, thus antisymmetric in
$a$ and $b$. As suggested in \cite {Kephart}, the potential can be
written in a more convenient form. It amounts to writing the vectors
in complex $3\times 3$ matrix notation. We can then interpret the
various terms in the scalar potential as invariant lie algebra
polynomials. Identifying
\begin{align}
H_1\sim (\bar{3},1,3)\rightarrow (q^c)_p^{\alpha} , \;\;H_2\sim
(3,\bar{3},1)\rightarrow (Q^a_{\alpha}), \;\; H_3\sim
(1,3,\bar{3})\rightarrow L_a^p\,,
\end{align}
the particle content of the MSSM in the above representation of the
model, is given by the expressions
\begin{eqnarray*}
 (\bar3,1,3)&\rightsquigarrow& q^c=\left(\begin{array}{ccc}
 d^r_R & u^r_R & D^r_R \\
 d^g_R & u^g_R & D^g_R \\
 d^b_R & u^b_R & D^b_R
 \end{array}
 \right)\;,\\
 (3,\bar 3,1)&\rightsquigarrow& Q=\left(\begin{array}{ccc}
 d^r_L & d^g_L & d^b_L \\
 u^r_L & u^g_L & u^b_L \\
 D^r_L & D^g_L & D^b_L
 \end{array}\right),\;\\
 (1,3,\bar3)&\rightsquigarrow& L=\left(\begin{array}{ccc}
 H_d^0 & H_u^+ & \nu_L \\
 H_d^- & H_u^0 & e_L \\
 \nu_R & e_R & S
 \end{array}\right)\;.
\end{eqnarray*}
Evidently $d_{L,R},u_{L,R},D_{L,R}$ transform as $3,\bar{3}$ under
color. Then we introduce
\begin{align*}
\hat{q}_{\alpha}^{c^p}=\frac{1}{3}\frac{\partial I_3}{\partial
q^{c\alpha}_p} , \;\; \hat{L}_{p}^a=\frac{1}{3}\frac{\partial
I_3}{\partial L^{p}_a}, \;\;
\hat{Q}^{\alpha}_a=\frac{1}{3}\frac{\partial I_3}{\partial
Q^a_{\alpha}} ,
\end{align*}
where $I_3$ is the trilinear $E_6 $ invariant term
$d_{ijk}A^{i}A^jA^k$ and $A$ is in $\mathbf{27}$ representation. The
$I_3$ term  can be decomposed under $SU(3)\times SU(3)\times SU(3)$
as\cite{Kephart}
\begin{equation}
I_3=det[Q]+det[q^c]+det[L]-tr(q^c\cdot L\cdot Q).
\end{equation}
In terms of these matrices, we have $\big< H_1|H_1\big>
=tr(q^{c\dagger}q^c),\big< H_2|H_2\big> =tr(Q^{\dagger}Q),\big<
H_3|H_3\big> =tr(L^{\dagger}L)$ and
\begin{align*}
d_{abc}H_1^aH_2^bH_3^c=detq^{c\dagger}+detM^{\dagger}+detL^{\dagger}-tr(N^{\dagger}M^{\dagger}L^{\dagger}).
\end{align*}
The $F$-terms which explicitly read
\begin{align*}
V_F=40d_{abc}d^{cde}(H_1^aH_2^bH_{1d}H_{2e}+H_2^aH_3^bH_{2d}H_{3e}+H_1^aH_3^bH_{1d}H_{3e}),
\end{align*}
can now be written as
\begin{align*}
V_F=40
tr(\hat{q}^{c^{\dagger}}\hat{q}^c+\hat{Q}^{\dagger}\hat{Q}+\hat{L}^{\dagger}\hat{L}).
\end{align*}
\subsection{Gauge symmetry breaking}
The spontaneous breaking of the $SU(3)_L$ and $SU(3)_R$ can be
triggered by the following vevs of the two families of $L$'s.
\[
\tilde L^3_0=\left(\begin{array}{ccc}
0 & 0&0\\
0&0&0\\
0&0&V_3
\end{array}\right),\;\;
\tilde L_0^2=\left(\begin{array}{ccc}
0 & 0&0\\
0&0&0\\
V_2&0&0
\end{array}\right).
\]
The action of $V_3$ breaks the gauge group according to
\begin{equation}
SU(3)_c\times SU(3)_L \times SU(3)_R\xrightarrow{V_3}SU(3)_c\times
SU(2)_L\times SU(2)_R\times U(1),
\end{equation}
while further breaking proceeds by  the action of $V_2$ to MSSM
\cite{Will}
\begin{equation}
SU(3)_c\times SU(2)_L\times SU(2)_R\times U(1)\xrightarrow{V_2}
SU(3)_c\times SU(2)_L\times U(1)_Y.
\end{equation}
Electroweak (EW) breaking then proceeds by the following MSSM vevs
\cite{MaZoup}
\[
\tilde L_0^1=\left(\begin{array}{ccc}
\upsilon_d& 0&0\\
0&\upsilon_u&0\\
0&0&0
\end{array}\right)\;.
\]
In the examination of the spontaneous symmetry breakings, the three
possible radii appearing in the potential are taken equal (strictly
speaking, this is the case that the manifold becomes nearly
K{\"a}hler).

It is worth noting that before the EW symmetry breaking,
supersymmetry is broken by both $D$-terms and $F$-terms, in addition
to its breaking by the soft terms. We plan to examine in detail the
phenomenological consequences of the resulting model, taking also
into account the massive Kaluza-Klein modes.

\section{Fuzzy spaces and fuzzy dimensional reduction}
\subsection{Geometry of the fuzzy sphere}

Let us build the discussion about fuzzy sphere \cite{madore} on the
familiar concept of the ordinary sphere, $S^2$. We may consider the
$S^2$ as a manifold embedded into the three dimensional Euclidean
space, $\mathbb{R}$. Therefore, the algebra of the functions on
$S^2$ can be specified through $\mathbb{R}^3$, by imposing the
constraint
\begin{equation}
  \sum_{a=1}^3x_a^2=R^2\,,\label{spherecondition}
\end{equation}
where $x_a$ are the coordinates in $\mathbb{R}^3$ and $R$ the radius
of the sphere. The isometry group of the sphere is a global $SO(3)$,
which is isomorphic to the $SU(2)$. Therefore, the generators of the
isometry group are $L_\alpha$, namely the three angular momentum
operators,
\begin{equation}
  L_a=-i\epsilon_{abc}x_b\pa_c\,.
\end{equation}
If we express the three operators $L_a$ in terms of the spherical
coordinates $\theta,\phi$, the above equation converts to
\begin{equation}
  L_a=-i\xi_a^\alpha\pa_\alpha\,,
\end{equation}
where the greek index, $\alpha$, denotes the spherical coordinates
and $\xi_a^\alpha$ are the components of the Killing vector fields
which generate the isometries of the sphere\footnote{The $S^2$
metric can be written in terms of the Killing vectors
$g^{\alpha\beta}=\dfrac{1}{R^2}\xi^\alpha_a\xi_a^\beta$.}.

Starting from the scalar Laplacian operator on the sphere,
$\triangle_{S^2}$
\begin{equation}
  \triangle_{S^2}=\frac{1}{\sqrt{g}}\pa_a(g^{ab}\sqrt{g}\pa_b)\,,
\end{equation}
one can find the spherical harmonics, $Y_{lm}(\theta,\phi)$, which
are the eigenfunctions of the $L^2$ operator
\begin{equation}
  L^2=-R^2\triangle_{S^2}\,.
\end{equation}
From the action of $L^2$ on the spherical harmonics, we also obtain
its eigenvalues
\begin{equation}
  L^2 Y_{lm}=l(l+1)Y_{lm}\,,
\end{equation}
where $l$ is non-negative integer. The spherical harmonics obey the
orthogonality condition
\begin{equation}
  \int \sin\theta d\theta d\phi Y_{lm}^\dag
  Y_{l'm'}=\delta_{ll'}\delta_{mm'}\,.
\end{equation}

Taking into consideration that spherical harmonics form a complete
and orthogonal set of functions, any function on $S^2$ can be
expanded in terms of $Y_{lm}(\theta,\phi)$
\begin{equation}
  a(\theta,\phi)=\sum_{l=0}^\infty\sum_{m=-l}^la_{lm}Y_{lm}(\theta,\phi)\,,\label{sphereexpansion}
\end{equation}
where $a_{lm}$ are complex coefficients. Besides the more frequent
expression of spherical harmonics in terms of spherical coordinates,
they can also be stated in terms of the cartesian coordinates, $x_a$
in $\mathbb{R}^3$,
\begin{equation}
  Y_{lm}(\theta,\phi)=\sum_{\vec{a}}f^{lm}_{a_1\ldots a_l}x^{a_1\ldots
  a_l}\,,\label{cartesianharmonics}
\end{equation}
where $f^{lm}_{a_1\ldots a_l}$ is a (traceless) symmetric tensor of
rank $l$.

Let us now describe the case of the fuzzy sphere in a comparative
way to the above description of the ordinary sphere. The fuzzy
sphere is the most typical case of non-commutative geometry, meaning
that the algebra of functions on a fuzzy sphere is not commutative
as it is on the ordinary sphere, due to the fact that it is
generated by spherical harmonics, with $l$ having a specific upper
limit. Therefore, instead of an infinite dimensional commutative
algebra, the algebra of a fuzzy sphere is truncated to finite
dimensional, in particular, $l^2$ dimensional, non-commutative
algebra. Thus, it is natural to consider the truncation of the
algebra as a matrix algebra and it is consistent to define the fuzzy
sphere as a matrix approximation of the ordinary sphere, $S^2$.

So, it follows that we are able to expand $N$-dimensional matrices
on a fuzzy sphere as
\begin{equation}
  \hat{a}=\sum_{l=0}^{N-1}\sum_{m=-l}^{l}a_{lm}\hat{Y}_{lm}\,,\label{fuzzyexpansion}
\end{equation}
where $\hat{Y}_{lm}$ are the fuzzy spherical harmonics, which are
now given by the expression
\begin{equation}
  \hat{Y}_{lm}=R^{-l}\sum_{vec{a}}f_{a_1\ldots
  a_l}^{lm}\hat{X}^{a_1}\cdots\hat{X}^{a_l}\,,
\end{equation}
where
\begin{equation}
  \hat{X}_a=\frac{2R}{\sqrt{N^2-1}}\lambda_a^{(N)}\,,\label{fuzzycoordinates}
\end{equation}
where $\lambda_a^{(N)}$ are the $SU(2)$ generators in the
$N$-dimensional representation and $f_{a_1\ldots a_l}^{lm}$ is the
same tensor that we met in \eqref{cartesianharmonics}. The fuzzy
spherical harmonics, $\hat{Y}_{lm}$ satisfy the orthonormality
condition
\begin{equation}
  \text{Tr}_N\left(\hat{Y}_{lm}^\dag\hat{Y}_{l'm'}\right)=\delta_{ll'}\delta_{mm'}\,.
\end{equation}

Moreover, there is a relation between the expansion of a function,
\eqref{sphereexpansion}, and that of a matrix,
\eqref{fuzzyexpansion} on the original and the fuzzy sphere,
respectively
\begin{equation}
  \hat{a}=\sum_{l=0}^{N-1}\sum_{m=-l}^{l}a_{lm}\hat{Y}_{lm}\quad\rightarrow\quad
  a=\sum_{l=0}^{N-1}\sum_{m=-l}^{l}a_{lm}Y_{lm}(\theta,\phi)\,.
\end{equation}
The above relation is obviously a map from matrices to functions.
Since we introduced the fuzzy sphere as a truncation of the algebra
of functions on $S^2$, considering the same $a_{lm}$ was just the
most natural choice. Of course, the choice of the map is not unique,
since it is not obligatory to consider the same expansion
coefficients $a_{lm}$.

Summing up the above analysis, the fuzzy sphere\cite{madore} is a
matrix approximation of the ordinary sphere, $S^2$. The cost we pay
for truncating the algebra of the functions is loss of
commutativity, so we end up with a non-commutative algebra, that of
matrices, $\text{Mat}(N;\mathbb{C})$. Therefore, fuzzy sphere,
$S_N$, is the non-commutative manifold with $\hat{X}_a$ being the
coordinate functions. As given by \eqref{fuzzycoordinates}, $X_a$
are $N\times N$ hermitian matrices which are generated be the
generators of $SU(2)$ in the $N-$dimensional representation. Of
course they have to respect both the condition
\begin{equation}
\sum_{a=1}^3\hat{X}_a\hat{X}_a=R^2\,,\label{fuzzyspherecondition}
\end{equation}
 which is the analogue of \eqref{spherecondition}, and the
commutation relations
\begin{equation}
[\hat{X}_a,\hat{X}_b]=i\alpha\epsilon_{abc}\hat{X}_c\,,\quad
  \alpha=\frac{2R}{\sqrt{N^2-1}}\,.\label{fuzzycommutationrelation}
\end{equation}

It is equivalent to consider the algebra on the fuzzy sphere being
described by the antihermitian matrices
\begin{equation}
  X_a=\frac{\hat{X}_a}{i\alpha R}\,,
\end{equation}
and satisfying the modified relations \eqref{fuzzyspherecondition},
\eqref{fuzzycommutationrelation}
\begin{equation}
  \sum_{a=1}^3X_aX_a=-\frac{1}{\alpha^2}\,,\quad
  [X_a,X_b]=C_{abc}X_c\,,
\end{equation}
where $C_{abc}=\dfrac{1}{R}\epsilon_{abc}$\,.

Let us proceed by stating the differential calculus on the fuzzy
sphere, which is three dimensional and obviously $SU(2)$ covariant.
Given a function $f$, its derivations along $X_a$ are
\begin{equation}
  e_a(f)=[X_a,f]\,,
\end{equation}
and consequently the Lie derivative on the function $f$ are
\begin{equation}
  \cl_af=[X_a,f]\,,\label{liederonfunction}
\end{equation}
where $\cl_a$ obey the Leibniz rule and the commutation relation of
$\mathfrak{su(2)}$
\begin{equation}
  [\cl_a,\cl_b]=C_{abc}\cl_c\,.\label{liecommutator}
\end{equation}
In the framework of differential forms, $\theta^a$ are the one-forms
dual to the vector fields $e_a$, namely
\begin{equation}
  \langle e_a,\theta^b\rangle=\delta^b_a\,.
\end{equation}
Therefore, the exterior derivative, $d$, of a function $f$ is given
by
\begin{equation}
  df=[X_a,f]\theta^a\,,
\end{equation}
as well as the action of the Lie derivative on the one-forms
$\theta^b$ is
\begin{equation}
  \cl_a\theta^b=C_{abc}\theta^c\,.\label{liederonform}
\end{equation}
Since the Lie derivative obeys the Leibniz law, its action on a
random one-form $\omega=\omega_a\theta^a$ gives
\begin{equation}
  \cl_b\omega=\cl_b(\omega_a\theta^a)=[X_b,\omega_a]\theta^a-\omega_aC^a_{bc}\theta^c\,,
\end{equation}
where we have applied \eqref{liederonfunction},
\eqref{liederonform}. Therefore, one obtains the result
\begin{equation}
  (\cl_b\omega)_a=[X_b,\omega_a]-\omega_cC^c_{ba}\,.
\end{equation}

After having detailed the differential geometry of the fuzzy sphere,
we are able to study the differential geometry of $M^4\times S^2_N$,
which is the product of Minkowski space and the fuzzy sphere with
$N-1$ fuzziness level. For instance, any one-form $A$ of $M^4\times
S^2_N$ can be expressed in terms of $M^4$ and the fuzzy sphere, that
is
\begin{equation}
  A=A_\mu dx^\mu+A_a\theta^a\,,\label{gaugepotentialoneform}
\end{equation}
where $A_\mu, A_a$ depend on both coordinates $x^\mu$ and $X_a$.

Furthermore,  instead of functions on the fuzzy sphere, one can
examine the case of spinors
\cite{aschieri-madore-manousselis-zoupanos}. Moreover, despite the
fact that for the present analysis we only include results about the
fuzzy sphere $S^2_N$, studying the differential geometry of other
higher-dimensional fuzzy spaces (e.g. fuzzy $CP^M$
\cite{balachandran, trivedi}) has been a task of great interest.

\subsection{Gauge theory on the fuzzy sphere}

In the previous subsection we studied how the algebra of functions
is modified when instead of the ordinary sphere, we consider the
fuzzy one. Given that we want to study gauge theory on the fuzzy
sphere, the next natural step is to to examine what happens if we
consider gauge fields on the fuzzy sphere. In order to do so, it is
prerequisite to introduce the covariant coordinates
\cite{madore-schrami-schup-wess}. So, at first, we consider a field
$\phi(X_a)$ on the fuzzy sphere, which depends on the powers of the
covariant coordinates, $X_a$ and then we take an infinitesimal
transformation, $\delta\phi$ of the gauge field
\begin{equation}
  \delta\phi(X)=\lambda(X)\phi(X)\,,
\end{equation}
where $\lambda(X)$ is the gauge transformation parameter. If
$\lambda(X)$ is an antihermitian function of $X_a$, the above gauge
transformation is infinitesimal Abelian $U(1)$. On the other hand,
if $\lambda(X)$ takes values in $\text{Lie}(U(P))$\footnote{We
assume that elements of $\text{Lie}(U(P))$ commute with the
covariant coordinates $X_a$.}, namely the algebra of hermitian
$P\times P$ matrices, then the above gauge transformation is
infinitesimal non-Abelian, $U(P)$. Also, it holds that
\begin{equation}
  \delta X_a=0\,,
\end{equation}
which ensures invariance of the covariant derivatives under a gauge
transformation. So, in the non-commutative case, left multiplication
by a coordinate is not a covariant operation, that is
\begin{equation}
  \delta(X_a\phi)=X_a\lambda(X)\phi\,,
\end{equation}
and generally it holds that
\begin{equation}
  X_a\lambda(X)\phi\neq\lambda(X)X_a\phi\,.
\end{equation}
One is then motivated by the non-fuzzy gauge theory to introduce the
covariant coordinates $\phi_a$, so that
\begin{equation}
  \delta(\phi_a\phi)=\lambda\phi_a\phi\,,
\end{equation}
which holds if
\begin{equation}
  \delta(\phi_a)=[\lambda,\phi_a]\,.\label{transformationofphia}
\end{equation}
The experience from the original (non-fuzzy) gauge theory also
dictates that
\begin{equation}
  \phi_a\equiv X_a+A_a\,,\label{covariantfield}
\end{equation}
with the $A_a$ being interpreted as the gauge potential of the
non-commutative theory. The equivalence is now obvious: $\phi_a$ is
the non-commutative analogue of the covariant derivative of ordinary
gauge theories. Combining \eqref{covariantfield} with
\eqref{transformationofphia} one is led to the transformation of
$A_a$, that is
\begin{equation}
  \delta A_a=-[X_a,\lambda]+[\lambda,A_a]\,,
\end{equation}
whose form encourages the interpretation of $A_a$ as a gauge field.
According to the above pattern of adjusting the content of the
original gauge theory to the fuzzy one, one proceeds with defining a
field strength tensor, $F_{ab}$, as
\begin{equation}
  F_{ab}\equiv[X_a,A_b]-[X_b,A_a]+[A_a,A_b]-C^c_{ab}A_c=[\phi_a,\phi_b]-C^c_{ab}\phi_c
\end{equation}
Although the last term of the above equation seems to spoil the
analogy, its imposition is necessary in order to preserve
covariance. In confirmation, it can be proved that the above field
strength tensor's transformation is covariant, that is
\begin{equation}
  \delta F_{ab}=[\lambda,F_{ab}]\,.
\end{equation}
\section{Ordinary fuzzy dimensional reduction}

Let us now perform an ordinary fuzzy dimensional reduction, applying
the structure we sketched above. We begin considering a
non-commutative gauge theory on the $M^4\times (S/R)_F$ space, with
gauge group $G=U(P)$, then we examine how it reduces to four
dimensions and finally give interpretations to the results. Let
$(S/R)_F$ be a fuzzy coset, e.g. the fuzzy sphere, $S^2_N$. The
action will be
\begin{equation}
  \mathcal{S}_{YM}=\frac{1}{4g^2}\int
  d^4xk\text{Tr}\text{tr}_GF_{MN}F^{MN}\,,\label{actiondimred}
\end{equation}
where $\text{tr}_G$ is the trace of the gauge group $G$ and
$k\text{Tr}$\footnote{Generally, $k$ is a parameter related to the
size of the fuzzy coset space. Especially in the case of the fuzzy
sphere, $k$ is a parameter which is related to the radius of the
sphere and the integer $l$.} denotes the integration over $(S/R)_F$,
that is the fuzzy coset which is described by $N\times N$ matrices.
$F_{MN}$ is the higher-dimensional field strength tensor, which
consists of  both four-dimensional spacetime and extra-dimensional
components, i.e. $(F_{\mu\nu},F_{\mu a},F_{ab})$.
The components of the field strength tensor, $F_{MN}$, in the extra
(non-commutative) directions, expressed in terms of the covariant
coordinates $\phi_a$ are
\begin{align}
  F_{\mu a}&=\pa_\mu\phi_a+[A_\mu,\phi_a]=D_\mu\phi_a\nonumber\\
  F_{ab}&=[X_a,A_b]-[X_b,A_a]+[A_a,A_b]-C_{ba}^cA_{ac}\,.\nonumber
\end{align}
Substituting the above equations in \eqref{actiondimred}, the action
takes the form
\begin{equation}
  \mathcal{S}_{YM}=\int
  d^4x\text{Tr}\text{tr}_G\left(\frac{k}{4g^2}F_{\mu\nu}^2+\frac{k}{2g^2}(D_\mu\phi_a)^2-V(\phi)\right)\,,\label{action2}
\end{equation}
where $V(\phi)$ denotes the potential term, which is the kinetic
term of $F_{ab}$, that is
\begin{align}
  V(\phi)&=-\frac{k}{4g^2}\text{Tr}\text{tr}_G\sum_{ab}F_{ab}F_{ab}\nonumber\\
  &=-\frac{k}{4g^2}\text{Tr}\text{tr}_G\left([\phi_a,\phi_b][\phi^a,\phi^b]-4C_{abc}\phi^a\phi^b\phi^c+2R^{-2}\phi^2\right)\,.
\end{align}
It is natural to interpret \eqref{action2} as an action in four
dimensions. Let $\lambda(x^\mu,X^a)$ be the gauge parameter that
appears in an infinitesimal gauge transformation of gauge group $G$.
This transformation can be considered as a $M^4$ gauge
transformation, so we write
\begin{equation}
  \lambda(x^\mu,X^a)=\lambda^I(x^\mu,X^a)\mathcal{T}^I=\lambda^{h,I}(x^\mu)T^h\mathcal{T}^I\,,\label{lambdareduction}
\end{equation}
where $\mathcal{T}^I$ denote the hermitian generators of the gauge
group $U(P)$, $\lambda^I(x^\mu,X^a)$ are $N\times N$ antihermitian
matrices, which means that they can be expressed as
$\lambda(x^\mu)^{I,h}T^h$, where $T^h$ are the antihermitian
generators of $U(N)$ and $\lambda(x^\mu)^{I,h}, h=1,\ldots, N^2$,
are the Kaluza-Klein modes of $\lambda(x^\mu,X^a)^I$. According to
this pattern, we assume that the fields on the right hand side of
\eqref{lambdareduction} may be considered as one field that takes
values in the tensor product Lie algebra
$\text{Lie}\left(U(N)\right)\otimes\text{Lie}\left(U(P)\right)$,
which is as a matter of fact the algebra
$\text{Lie}\left(U(NP)\right)$. Moreover, the gauge field $A_a$ can
be written as
\begin{equation}
  A_\nu(x^\mu,X^a)=A_\nu^I(x^\mu,X^a)\mathcal{T}^I=A_\nu^{h,I}T^h\mathcal{T}^I\,,
\end{equation}
which is similarly interpreted as a gauge field on $M^4$ that is
valued in the $\text{Lie}\left(U(NP)\right)$ algebra. Similar
interpretation can be given in the case of the scalar fields, too.
It is worth-noting that the scalars are accommodated in the adjoint
representation of the four-dimensional gauge group, which means that
they are incapable of inducing the electroweak symmetry breaking.
This fact is a significant motivation to proceed to non-trivial
dimensional reduction schemes, like the one we shall present in the
following section. At last, $\text{Tr}\text{tr}_G$ is interpreted as
the trace of the $U(NP)$ matrices.

Before we proceed to the next (non-trivial) dimensional reduction
scheme, we may state two important remarks on the previous
(ordinary) one. First, due to non-commutativity, the space of
functions is finite-dimensional, which leads to the fact that the
tower of modes is a finite sum, given by the trace $\text{Tr}$ and
second, we observe that after the above fuzzy dimensional reduction,
we have resulted in a four-dimensional gauge theory whose group is
enhanced compared to the initial gauge group, $G$ of the
higher-dimensional theory. Thus, we deduce that the initial gauge
group $G$ is not necessarily non-Abelian in order to result with
such one in four-dimensions. An Abelian gauge group in the
higher-dimensional theory should be a valid choice.

\section{Fuzzy CSDR}
In the previous section we applied an ordinary dimensional reduction
but the resulting four-dimensional theory was defective. Therefore,
in order to obtain a more appropriate gauge theory in four
dimensions, we proceed to apply a non-trivial dimensional reduction,
i.e. the fuzzy extension of the Coset Space Dimensional Reduction
(CSDR).

So, in this section we adjust the CSDR programme -described
explicitly in first part- in the non-commutative case, where the
extra dimensions are fuzzy coset spaces
\cite{aschieri-madore-manousselis-zoupanos}\footnote{See also
\cite{Harland-kurkcuoglu}.}, in order to achieve a reduction of the
number of both gauge and scalar fields in the action
\eqref{action2}. In general, the group $S$ acts on the fuzzy coset
$(S/R)_F$, and as we have seen in the commutative case, CSDR scheme
suggests that the fields of the theory must be invariant under an
infinitesimal group transformation of $S$, up to an infinitesimal
gauge transformation. Specifically, the fuzzy coset we use is the
fuzzy sphere, $(SU(2)/U(1))_F$, so the action of an infinitesimal
$SU(2)$ transformation on the fuzzy sphere should leave the scalar
and gauge fields invariant, up to an infinitesimal gauge
transformation
\begin{align}
  \cl_b\phi&=\delta_{W_b}=W_b\phi \label{csdrconstraintone} \\
  \cl_bA&=\delta_{W_b}A=-DW_b\,, \label{csdrconstrainttwo}
\end{align}
where $A$ is the gauge potential expressed as an one-form, see
\eqref{gaugepotentialoneform} and $W_b$ is a gauge parameter that it
is antihermitian and depends only on the coset coordinates $X^a$.
Therefore, we can write $W_b$ as
\begin{equation}
  W_b=W_b^I\mathcal{T}^I\,,\quad I=1,2,\ldots, P^2\,,
\end{equation}
where $\mathcal{T}^I$ are hermitian generators of $U(P)$ and
$(W_b^I)^\dag=-W_b^I$, where the $^\dag$ denotes the hermitian
conjugation on the $X^a$ coordinates.

Using the covariant coordinates $\phi_a$, \eqref{covariantfield},
and $\omega_a$ which is defined as
\begin{equation}
  \omega_a\equiv X_a-W_a\,,
\end{equation}
the CSDR constraints, \eqref{csdrconstraintone},
\eqref{csdrconstrainttwo}, take the following, rather simple, form
\begin{align}
  [\omega_b,A_\mu]&=0\label{constraintone}\\
  C_{bde}\phi^e&=[\omega_b,\phi_d]\label{constrainedtwo}\,.
\end{align}
As we have mentioned before, \eqref{liecommutator}, due to Lie
derivatives respect the $\mathfrak{su(2)}$ commutation relation, one
ends up with the following consistency condition
\begin{equation}
  [\omega_a,\omega_b]=C_{ab}^c\omega_c\,,\label{consistencycondition}
\end{equation}
where the transformation of $\omega_a$ is
\begin{equation}
  \omega_a\,\rightarrow\,\omega_a'=g\omega_ag^{-1}\,.
\end{equation}
In the case of spinor fields, the procedure is quite similar
\cite{aschieri-madore-manousselis-zoupanos}.

\subsection{Solving the CSDR constraints for the fuzzy sphere}

Let us now study the basic example of fuzzy CSDR, in which we
consider the fuzzy coset to be the fuzzy sphere, $(S/R)_F=S_N^2$ and
the gauge group to be $G=U(1)$. The $\omega_a=\omega_a(X^b)$ are
$N\times N$ antihermitian matrices, which means that they can be
regarded as elements of $\text{Lie}(U(N))$, but also they satisfy
the commutation relation of $\text{Lie}(SU(2))$, as given in the
consistency condition, \eqref{consistencycondition}. So, since we
require the consistency condition, \eqref{consistencycondition}, to
hold, we have to embed $\text{Lie}(SU(2))$ into $\text{Lie}(U(1))$.
Therefore, if $T^h,\, h=1,\ldots,N^2$ are the $\text{Lie}(U(N))$
generators, in the fundamental representation, then we can always
use the convention $h=(a,u),\,a=1,2,3\,,\,u=4,5,\ldots,N^2$, with
the generators $T^a$ satisfying the $\text{Lie}(SU(2))$
\begin{equation}
  [T^a,T^b]=C^{ab}_cT^c\,.\label{relationtorespect}
\end{equation}
and the embedding is finally defined by identifying
\begin{equation}
  \omega_a=T_a\,.
\end{equation}
So, the constraint \eqref{constraintone} suggests that the gauge
group $K$ of the four-dimensional theory is the centralizer of the
image of $SU(2)$ into $U(N)$, that is
\begin{equation}
  K=C_{U(N)}(SU(2))=SU(N-2)\times U(1)\times U(1)\,,
\end{equation}
where the second $U(1)$ in the right hand side is the one that comes
from
\begin{equation}
  U(N)\simeq SU(N)\times U(1)\,.
\end{equation}
Therefore, as for $x$, the $A_\mu(x,X)$ are arbitrary functions, but
on the other hand they depend on $X$ in a way that takes values in
$\text{Lie}(K)$ instead of $\text{Lie}(U(N))$. To sum up, that means
that we end up with a four-dimensional gauge potential valued in
$\text{Lie}(K)$.

Let us now examine the second constraint, \eqref{constrainedtwo}.
This one is satisfied if we choose
\begin{equation}
  \phi_a=r\phi(x)\omega_a\,,
\end{equation}
which means that the degrees of freedom which remain unconstrained
are related to the scalar field $\phi(x)$, which is belongs to the
trivial representation of the four-dimensional gauge group $K$
(singlet under $K$).

In the above example the need of embedding $SU(2)$ into $U(N)$ was
emerged by the consistency condition \eqref{consistencycondition}.
Although the embedding was realized into the fundamental
representation of $U(N)$, we could have used the irreducible
$N$-dimensional representation of $SU(2)$ by identifying
$\omega_a=X_a$. In this case, the constraint \eqref{constraintone}
leads to the fact that $U(1)$ is the four-dimensional gauge group,
so that $A_\mu(x)$ is valued in $U(1)$. The second constrained,
\eqref{constrainedtwo}, implies that, in this case too, $\phi(x)$ is
a scalar singlet.

To sum up the above procedure, one starts with a $U(1)$ gauge theory
on $M^4\times S^2_N$ and due to the consistency condition,
\eqref{consistencycondition}, an embedding of $SU(2)$ into $U(N)$ is
required\footnote{This embedding can be achieved in several ways,
more specifically in $p_N$ ways, where $p_N$ is the possible ways
one can partition the $N$ into a set of non-increasing, positive
integers \cite{madore}.}. Then, the first CSDR constraint,
\eqref{constraintone}, produces the four-dimensional gauge group and
the second one, \eqref{constrainedtwo}, produces the
four-dimensional scalar fields that survive from the reduction.

As far as the fermions are concerned, we are going to list briefly
the results of the above procedure. According to the extended
analysis \cite{aschieri-madore-manousselis-zoupanos}, it proves that
the suitable embedding is
\begin{equation}
  S\subset SO(\text{dim}S)\,,
\end{equation}
which is achieved by $T_a=\dfrac{1}{2}C_{abc}\Gamma^{bc}$, which
respects \eqref{relationtorespect}. Thus, $\psi$ is an interwining
operator between the representations of $S$ and $SO(\text{dim}S)$.
In accordance to the commutative case \cite{kapetanakis-zoupanos},
in order to find the surviving fermions in four-dimensional theory,
we have to decompose the adjoint representation of $U(NP)$ under the
product $S_{U(NP)}\times K$, that is
\begin{align}
  U(NP)\supset &\,S_{U(NP)}\times K\,,\\
 \text{adj}\,U(NP)&=\sum_i(s_i,k_i)\,.
\end{align}
Also, the decomposition of the spinorial representation $\sigma$ of
$SO(\text{dim}S)$ under $S$
\begin{align}
  SO&(\text{dim}S)\supset S\,,\\
  \sigma&=\sum_e\sigma_e\,.
\end{align}
Therefore, in the case where the two irreducible representations
$s_i, \sigma_e$ are identical, the surviving fermions
(four-dimensional spinors) of the four-dimensional theory belong to
the $k_i$ representation of gauge group $K$.

Before we move on, this is an appropriate point to compare the
ordinary higher-dimensional theory $M^4\times (S/R)$, to its fuzzy
extension $M^4\times (S/R)_F$. We begin with the similarities:
fuzziness does not affect the isometries - both spaces have the
same, $SO(1,3)\times SO(3)$ and the gauge couplings defined on the
two spaces have the same dimensionality. On the other hand there is
a very striking difference: of the two, only the non-commutative
higher-dimensional theory is renormalizable\footnote{Meaning that
the number of counter-terms required to eliminate the divergencies
is finite.}. In addition, a $U(1)$, defined on the $M^4\times
(S/R)_F$ space, is enough in order to end up with non-abelian
structures in four dimensions\footnote{Technically, this is possible
because $N\times N$ matrices can be decomposed on the $U(N)$
generators.}.

\section{Orbifolds and fuzzy extra dimensions}

The need for chiral low energy theories in the framework of gauge
theories with fuzzy extra dimensions has motivated the introduction
of the orbifold structure, similar to the one developed in
\cite{kachru-silverstein}. This technique is an alternative option
to the standard one for obtaining $\mathcal{N}=1$ four dimensional
models, that is reducing the theory on suitable manifolds, e.g.
Calabi-Yau manifolds \cite{candelas-horowitz} or manifolds with
$SU(3)$-structure (see \cite{cardoso-curio, gauntlet-martelli}). The
authors of \cite{kachru-silverstein} were motivated by the duality
between four-dimensional $\mathcal{N}=4$, $U(N)$ SYM theory and Type
$IIB$ string theory on $AdS_5\times S^5$ \cite{maldacena}, so they
applied orbifold techniques similar to \cite{douglas,
douglas-greene-morrison} in order to break some of the four
supersymmetries. Also, the starting gauge group, $SU(3N)$, which is
realized on $3N$ $D3$ branes\footnote{This is the meeting point of
the two different frameworks (superstring theories and
non-commutative geometry) that aim at unification, i.e.
non-commutative gauge theory can describe effective physics on
D-branes.}, breaks down to $SU(N)^3$ with the fermions being nested
into chiral representations of the latter gauge group.

So, the concept of deconstruction of dimensions \cite{arkani},
motivated the idea to reverse the above
\cite{aschieri-grammatikopoulos, steinacker-zoupanos,
chatzistavrakidis-steinacker-zoupanos} in order to further justify
the renormalizability of the theory and to attempt the building of
chiral models in theories arising from the framework of fuzzy extra
dimensions. Moreover, the reversed procedure gives hope that it is
not necessary to consider the initial abelian gauge theory as
higher-dimensional, but the non-abelian gauge theory can emerge from
fluctuations of the coordinates \cite{steinacker1}. This
consideration is realized as follows: one starts with a
four-dimensional gauge theory, includes an appropriate scalar
spectrum and a suitable potential which can lead to vacua that could
be interpreted as dynamically generated fuzzy extra dimensions and
they also include a finite Kaluza-Klein tower of massive modes.

Of course it was desired to include fermions in such models but the
best one had achieved so far, was to obtain theories with mirror
fermions in bifundamental representations of the low-energy gauge
group
\cite{steinacker-zoupanos,chatzistavrakidis-steinacker-zoupanos}.
Mirror fermions do not exclude the possibility to make contact with
phenomenology \cite{maalampi-roos}, nevertheless, it is preferred to
end up with exactly chiral fermions.

In this section the plan that was sketched above is realized. A
dimensional reduction on an orbifold \cite{dixon-harvey,
bailin-love} is performed in order to achieve $\mathcal{N}=1$
supersymmetric chiral theories in four dimensions. Specifically, in
this review, we are going to deal with the $\mathbb{Z}_3$ orbifold
projection of the $\mathcal{N}=4$ Supersymmetric Yang Mills (SYM)
theory \cite{brink-schwarz-scherk}, examining the action of the
discrete group on the theory's fields and the superpotential that
emerges in the projected theory.

\subsection{$\mathcal{N}=4$ SYM field theory and $\mathbb{Z}_3$ orbifolds \label{subsec:subsection}}
So, let us begin with a $\mathcal{N}=4$ supersymmetric $SU(3N)$
gauge theory. The particle spectrum of the theory (in the
$\mathcal{N}=1$ terminology) consists of an $SU(3N)$ gauge
supermultiplet, three adjoint chiral supermultiplets
$\Phi^i\,,i=1,2,3$. The component fields of the above
supermultiplets are the gauge bosons, $A_\mu,\,\mu=1,\ldots,4$, six
adjoint real (or three complex) scalars $\phi^a,\,a=1,\ldots,6$ and
four adjoint Weyl fermions $\psi^p,\,p=1,\ldots,4$. The scalars and
Weyl fermions transform under the $6$ and $4$ representation of the
$SU(4)_R$ $R$-symmetry of the theory, respectively. The whole theory
is of course defined on the Minkowski spacetime.

Then, in order to introduce orbifolds, the discrete group
$\mathbb{Z}_3$ has to be considered as a subgroup of $SU(4)_R$. The
embedding of $\mathbb{Z}_3$ into $SU(4)_R$ is not unique, but the
options are not equivalent, since the choice of embedding
differentiates the amount of remnant supersymmetry
\cite{kachru-silverstein}:
\begin{enumerate}
  \item Maximal embedding of $\mathbb{Z}_3$ into $SU(4)_R$ would
  lead to non-supersymmetric models, therefore this choice is
  excluded
  \item Embedding of $\mathbb{Z}_3$ in an $SU(4)_R$ subgroup:
  \begin{itemize}
    \item[-] Embedding into an $SU(2)$ subgroup would lead to
    $\mathcal{N}=2$ supersymmetric models with $SU(2)_R$ $R$-symmetry
    \item[-] Embedding into an $SU(3)$ subgroup would lead to
    $\mathcal{N}=1$ supersymmetric models with $U(1)_R$
    $R$-symmetry.
  \end{itemize}
\end{enumerate}

 We shall focus on the last option which is the desired one, since it leads to
$\mathcal{N}=1$ supersymmetric models. Let us consider a generator
$g\in\mathbb{Z}_3$, which is (for convenience) labeled by three
integers $\vec{a}=(a_1,a_2,a_3)$ \cite{douglas-greene-morrison}
which satisfy the relation
\begin{equation}
  a_1+a_2+a_3=0\,\,\text{mod}\,3\,.
\end{equation}
The last equation implies the fact that $\mathbb{Z}_3$ is embedded
in $SU(3)$, i.e. the fact that the remnant supersymmetry is
$\mathcal{N}=1$ \cite{bailin-love}.

It is expected that since the various fields of the theory transform
differently under $SU(4)_R$, $\mathbb{Z}_3$ will generally act
non-trivially on them:
\begin{itemize}
  \item Gauge and gaugino fields are singlets under $SU(4)_R$,
  therefore the geometric action of the $\mathbb{Z}_3$ rotation is
  trivial.
  \item The action of $\mathbb{Z}_3$ on the complex scalar fields is
  given by the matrix
  \begin{equation}
    \gamma(g)_{ij}=\delta_{ij}\omega^{a_i}\,,
  \end{equation}
  where $\omega=e^{\frac{2\pi}{3}}$\,.
  \item The action of $\mathbb{Z}_3$ on the fermions $\phi^i$ is
  given by
  \begin{equation}
    \gamma(g)_{ij}=\delta_{ij}\omega^{b_i}\,,
  \end{equation}
  where $b_i=-\dfrac{1}{2}(a_{i+1}+a_{i+2}-a_i)$\footnote{Also modulo 3}.
\end{itemize}
In the case under study the three integers of the generator $g$ are
$(1,1,-2)$, which means that $a_i=b_i$.

The fact that the matter fields are not invariant under a gauge
transformation, $\mathbb{Z}_3$ acts on their gauge indices, too. The
action of this rotation is specified by the matrix
\begin{equation}
  \gamma_3=\left(
             \begin{array}{ccc}
               \mathbf{1}_N & 0 & 0 \\
               0 & \omega\mathbf{1}_N & 0 \\
               0 & 0 & \omega^2\mathbf{1}_N \\
             \end{array}
           \right)\,.\label{gammatria}
\end{equation}

A priori, there is no reason why these blocks should have the same
dimensionality (see e.g.\cite{aldabaz-ibanez,
lawrence-nekrason-vafa, kiritsis}). However, since the projected
theory must be anomaly free, the dimension of the three blocks is
the same.

After the orbifold projection, the derived theory must contain the
fields that are invariant under the combined action of the discrete
group, $\mathbb{Z}_3$, on the "geometric"\footnote{In case of
ordinary reduction of a $10$-dimensional $\mathcal{N}=1$
supersymmetric Yang-Mills theory, one obtains an $\mathcal{N}=4$
supersymmetric Yang-Mills theory in four dimensions. This has a
global $SU(4)_R$ symmetry which is identified with the tangent space
$SO(6)$ of the extra dimensions \cite{manousselis-zoupanos,
manousselis-zoupanos2}.} and gauge indices
\cite{douglas-greene-morrison}. As far as the gauge bosons is
concerned, the projection is
\begin{equation}
  A_\mu=\gamma_3 A_\mu\gamma_3^{-1}\,.
\end{equation}
So, taking into consideration \eqref{gammatria}, the gauge group of
the initial theory breaks down to the group $H=SU(N)\times
SU(N)\times SU(N)$ in the projected theory.

As we have already mentioned, the complex scalar fields transform
non-trivially both under the gauge and $R-$symmetry, so the
projection is
\begin{equation}
  \phi_{IJ}^i=\omega^{I-J+a_i}\phi^i_{IJ}\,,
\end{equation}
where $I,J$ are gauge indices. Therefore, $J=I+a_i$, which means
that the scalar fields that survive the orbifold projection have the
form $\phi_{I,J+a_i}$ and their transformation under the gauge group
$H$ is
\begin{equation}
  3\cdot\left((N,\bar{N},1)+(\bar{N},1,N)+(1,N,\bar{N})\right)\,.\label{repaftrprojection}
\end{equation}

Similarly, fermions transform non-trivially under the gauge group
and $R-$symmetry, too. Therefore, the projection is
\begin{equation}
  \psi^i_{IJ}=\omega^{I-J+b_i}\psi_{IJ}^i\,.
\end{equation}
Therefore, the fermions that survive the orbifold projection have
the form $\psi^i_{I,I+b_i}$ and they belong to the same
representation of $H$ as the scalars, that is
\eqref{repaftrprojection}. The fact that scalars and fermions share
common representations, demonstrates the $\mathcal{N}=1$ remnant
supersymmetry. It is worth noting that after the orbifold
projection, the representations \eqref{repaftrprojection} of the
resulting theory are anomaly free. If we had not taken into account
the requirement that the three blocks of \eqref{gammatria} had to be
of the same dimensionality, the resulting theory would not be
anomaly free. Therefore, the need of additional sectors would emerge
in order to achieve cancelation of the anomalies.

Let us now make a few comments about the above results as far as the
fermions is concerned. First, the fermions are accommodated in
chiral representations of $H$ and second, there are three fermionic
generations since, as we have mentioned above, the particle spectrum
contains three $\mathcal{N}=1$ chiral supermultiplets.

It is known that the interactions of a supersymmetric model are
given by the superpotential. In order to specify the superpotential
of the projected theory, it is necessary to begin with the
superpotential of the original $\mathcal{N}=4$ SYM theory
\cite{brink-schwarz-scherk}
\begin{equation}
  W_{\mathcal{N}=4}=\epsilon_{ijk}\text{Tr}(\Phi^i\Phi^j\Phi^k)\,,
\end{equation}
where, $\Phi^i,\Phi^j,\Phi^k$ refer to the three chiral superfields
of the theory. After the projection the structure of the
superpotential does not change, but it encrypts only the
interactions among the surviving fields of the $\mathcal{N}=1$
theory, that is
\begin{equation}
  W_{\mathcal{N}=1}^{(proj)}=\sum_I\epsilon_{ijk}\Phi_{I,I+a_i}^i\Phi_{I+a_i,I+a_i+a_j}^j\Phi_{I+a_i+a_j,I}^k\,.\label{projectedsuper}
\end{equation}

\subsection{Dynamical generation of twisted fuzzy
spheres}

As we have already mentioned, the superpotential of the projected
$\mathcal{N}=1$ theory maintains the form of the initial
superpotential $\mathcal{N}=4$, but contains only the field that
passed through the orbifold filtering. From the superpotential
$W_{\mathcal{N}=1}^{proj}$ that is given in \eqref{projectedsuper},
the scalar potential can be extracted:
\begin{equation}
  V_{\mathcal{N}=1}^{proj}(\phi)=\frac{1}{4}\text{Tr}\left([\phi^i,\phi^j]^\dag[\phi^i,\phi^j]\right)\,,\label{scalarpotential}
\end{equation}
where, obviously, $\phi^i$ are the scalar component fields of the
superfield $\Phi^i$. The above potential is minimized by vanishing
vevs of the fields, so modifications have to be made so that
solutions that can be interpreted as vacua of a non-commutative
geometry to be emerged.

So, in order to result with a minimum of the scalar potential,
$\mathcal{N}=1$ soft supersymmetric terms of the form\footnote{The
SSB terms that will be inserted into the scalar potential, are
purely scalar. Although this satisfies our purpose, it is obvious
that other SSB terms have to be included too in order to obtain the
full SSB sector \cite{djouadi}.}
\begin{equation}
  V_{SSB}=\frac{1}{2}\sum_im_i^2\phi^{i\dag}\phi^i+\frac{1}{2}\sum_{i,j,k}h_{ijk}\phi^i\phi^j\phi^k+h.c.\label{ssbterms}
\end{equation}
are introduced, where $h_{ijk}=0$ unless
$i+j+k\equiv0\,\text{mod}3$. The introduction of the above SSB terms
should not come as a surprise, since an SSB sector is necessary
anyway for a model with realistic aspirations, see
e.g.\cite{djouadi}. It is also necessary to include the $D$-terms of
the theory, which are given by
\begin{equation}
  V_D=\frac{1}{2}D^2=\frac{1}{2}D^ID_I\,,
\end{equation}
where $D^I=\phi_i^\dag T^I\phi^i$, where $T^I$ are the generators of
the representation of the corresponding chiral multiplets.

So, after the introduction of the SSB terms \eqref{ssbterms} in the
scalar potential \eqref{scalarpotential}, as well as the inclusion
of $D$-terms, the total potential of the theory now is
\begin{equation}
V=V_{\mathcal{N}=1}^{proj}+V_{SSB}+V_D\,.\label{totalscalarpotential}
\end{equation}
A suitable choice for $m_i^2$ and $h_{ijk}$ parameters in
\eqref{ssbterms} is
\begin{equation}
  m_i^2=1\,,\quad h_{ijk}=\epsilon_{ijk}\,.
\end{equation}
Therefore, the total scalar potential, \eqref{totalscalarpotential},
takes the form
\begin{equation}
  V=\frac{1}{4}(F^{ij})^\dag F^{ij}+V_D\,,\label{twistedpotential}
\end{equation}
where $F^{ij}$ is defined as
\begin{equation}
  F^{ij}=[\phi^i,\phi^j]-i\epsilon^{ijk}(\phi^k)^\dag\,.\label{fuzzyfieldstrength}
\end{equation}

It is obvious that the first term of the scalar potential,
\eqref{twistedpotential}, is always positive. Therefore, the global
minimum of the potential is obtained when
\begin{equation}
  [\phi^i,\phi^j]=i\epsilon_{ijk}(\phi^k)^\dag\,,\quad
  \phi^i(\phi^j)^\dag=R^2\,,\label{twistedfuzzy}
\end{equation}
where $(\phi^i)^\dag$ denotes the hermitian conjugate of the complex
scalar field $\phi^i$ and $[R^2,\phi^i]=0$. It is clear that the
above equations are related to a fuzzy sphere. This may arise by
considering the untwisted fields $\tilde{\phi}^i$, defined by
\begin{equation}
  \phi^i=\Omega\tilde{\phi}^i\,,\label{twisted-untwisted}
\end{equation}
where $\Omega\neq1$ satisfying the relations
\begin{align}
\Omega^3=1\,,\,\,\,\,[\Omega,\phi^i]=0\,,\,\,\,\,\Omega^\dag=\Omega^{-1}\,,\,\,\,\,
(\tilde{\phi}^i)^\dag=\tilde{\phi}^i\,\,\Leftrightarrow\,\,
(\phi^i)^\dag=\Omega\phi^i\,.
\end{align}
Therefore, \eqref{twistedfuzzy} reproduces the ordinary fuzzy sphere
relations generated by $\tilde{\phi}^i$
\begin{equation}
  [\tilde{\phi}^i,\tilde{\phi}^j]=i\epsilon_{ijk}\tilde{\phi}^k\,,\quad\tilde{\phi}^i\tilde{\phi}^i=R^2\,,\label{untwisted}
\end{equation}
exhibiting the reason why the non-commutative space generated by
$\phi^i$ is called a twisted fuzzy sphere, $\tilde{S}_N^2$.
Remarkably, the above structure is valid only for $\mathbb{Z}_3$,
excluding the choice of another cyclic group $\mathbb{Z}_n$ as the
orbifold group.

It is now straightforward to find configurations of the twisted
fields $\phi^i$, i.e. fields that satisfy the relations
\eqref{twistedfuzzy}. Such a configurations is
\begin{equation}
  \phi^i=\Omega(\mathbf{1}_3\otimes\lambda^i_{(N)})\,,
\end{equation}
where $\lambda^i_{(N)}$ are the $SU(2)$ generators in the
$N$-dimensional irreducible representation and $\Omega$ is the
matrix given by
\begin{equation}
  \Omega=\Omega_3\otimes\mathbf{1}_N\,,\quad\Omega_3=\left(
                                                       \begin{array}{ccc}
                                                         0 & 1 & 0 \\
                                                         0 & 0 & 1 \\
                                                         1 & 0 & 0 \\
                                                       \end{array}
                                                     \right)\,,\quad\Omega^3=\mathbf{1}\,.
\end{equation}
According to the transformation \eqref{twisted-untwisted}, the
"off-diagonal" orbifold sectors \eqref{repaftrprojection} take the
following block-diagonal form
\begin{equation}
  \phi^i=\left(
           \begin{array}{ccc}
             0 & (\lambda_{(N)}^i)_{(N,\bar{N},1)} & 0 \\
             0 & 0 & (\lambda_{(N)}^i)_{(1,N,\bar{N})} \\
             (\lambda_{(N)}^i)_{(\bar{N},1,N)} & 0 & 0 \\
           \end{array}
         \right)=\Omega\left(
                         \begin{array}{ccc}
                           \lambda^i_{(N)} & 0 & 0 \\
                           0 & \lambda^i_{(N)} & 0 \\
                           0 & 0 & \lambda^i_{(N)} \\
                         \end{array}
                       \right)\,.\label{triaseena}
\end{equation}
So, it is understood that the untwisted fields that generate the
ordinary fuzzy sphere, $\tilde{\phi}^i$, are expressed in a
block-diagonal form. Each block can be regarded as an ordinary fuzzy
sphere, since they separately satisfy the corresponding commutation
relations \eqref{untwisted}. Therefore, the above configuration
\eqref{triaseena} can be interpreted as three fuzzy spheres
(branes), embedded with relative angles $2\pi/3$. In conclusion, the
solution $\phi^i$ can be regarded as the twisted equivalent of three
fuzzy spheres which conform with the orbifolding.

Summing up, we can say that, at least for a fixed range of
parameters, the global minimum of the scalar potential is achieved
by a twisted fuzzy sphere, $\tilde{S}^2_N$. So, the expression
$F^{ij}$ that was defined above in \eqref{fuzzyfieldstrength}, will
be interpreted as the field strength on the spontaneously generated
fuzzy extra dimensions. The second term of the potential, $V_D$,
settles for a change on the radius of the sphere, similar to the one
in the case of the ordinary fuzzy sphere
\cite{aschieri-grammatikopoulos, steinacker,
chatzistavrakidis-steinacker-zoupanos}.

Let us now focus on the geometric point of view of the potential's
vacuum. As we have already discussed, the scalar component fields of
the theory, $\phi^i$, form the potential,
\eqref{totalscalarpotential}, through $F$-terms, $D$-terms and SSB
terms. Fixing the parameters leads to minimization of the potential
by a twisted fuzzy sphere solution
\begin{equation}
  \phi^i=\Omega(\mathbf{1}_3\otimes(\lambda^i_{(N-n)})\oplus 0_n)\,,
\end{equation}
where $0_n$ is the $n\times n$ matrix with zero entries. This
non-vanishing vacuum, which should be regarded as
$\mathbb{R}^4\times\tilde{S}_N^2$ with a twisted fuzzy sphere in
$\phi^i$ coordinates, leads to the breaking of the gauge symmetry,
$SU(N)^3$ down to $SU(n)^3$.

Now, the fluctuations of the scalar fields around the vacuum can be
understood by considering the transformation
\eqref{twisted-untwisted}, $\phi^i=\Omega\tilde{\phi}^i$. In the
non-twisted case, fluctuations around the vacuum describe scalar and
gauge fields on $S^2_N$ \cite{madore-schrami-schup-wess,
steinacker1}, which gain mass from the $\mathbb{R}^4$ point of view.
The \eqref{triaseena} informs us that the matrix $\Omega$ maps the
twisted fuzzy sphere into three ordinary fuzzy spheres as three
$N\times N$ blocks are diagonally embedded into the original
$3N\times 3N$ matrix. Therefore, all fluctuations could be regarded
as fields on the three untwisted fuzzy spheres
\begin{equation}
  \phi^i=\tilde{\Omega}(\lambda^i_{(N)}+A^i)=\left(\begin{array}{ccc}
                                               \lambda^i_{(N)}+A^i & 0 & 0 \\
                                               0 & \omega(\lambda^i_{(N)}+A^i) & 0 \\
                                               0 & 0 & \omega^2(\lambda^i_{(N)}+A^i)
                                             \end{array}\right)\,,
\end{equation}
as well as specific massive off-diagonal states which cyclically
connect these spheres. The field strength $F^{ij}$,
\eqref{fuzzyfieldstrength}, converts to
\begin{equation}
  F^{ij}=[\phi^i,\phi^j]-i\epsilon^{ijk}(\phi^k)^\dag=\Omega^2\left([\tilde{\phi}^i,\tilde{\phi}^j]
  -i\epsilon^{ijk}\tilde{\phi}^k\right)\,,
\end{equation}
that is the field strength on an untwisted fuzzy sphere. Thus, at
intermediate energy scales, the vacuum can be interpreted as
$\mathbb{R}^4\times S_N^2$ with three untwisted fuzzy spheres in the
$\tilde{\phi}^i$ coordinates. The three spheres are not mixed due to
the lack of off-diagonal entries, due to the orbifold projection. As
in \cite{aschieri-grammatikopoulos, steinacker-zoupanos,
chatzistavrakidis-steinacker-zoupanos}, because of the Higgs effect,
gauge fields and fermions decompose to a finite Kaluza-Klein tower
of massive modes on $S^2_N$ resp. $\tilde{S}^2_N$, as well as a
massless sector.

\section{Chiral models after the fuzzy orbifold projection}

Let us now sum up the above context, then note the possible valid
models that emerge from the initial framework and finally focus on
the most interesting one.

For all cases the initial theory is common, that is the
$\mathcal{N}=4$ SYM four-dimensional theory that is governed by
$SU(3N)$ gauge symmetry. We have already listed the field spectrum,
which is an $SU(3N)$ gauge supermultiplet an three adjoint chiral
supermultiplets $\Phi^i$. The superpotential that encodes the
interactions of the model is
\begin{equation}
  W_{\mathcal{N}=4}=\epsilon_{ijk}\text{Tr}(\Phi^i\Phi^j\Phi^k)\,.
\end{equation}
Then, it follows the choice of a suitable discrete group, i.e.
$\mathbb{Z}_3$, which is embedded into the $SU(3)$ subgroup of
$SU(4)_R$ in order to realize the orbifold projection of the theory.
After the projection, the $SU(3N)$ gauge group is broken to the
$\mathcal{N}=1$ $SU(N)^3$ theory with the scalars and fermions
nested into chiral representations of the gauge group. Specifically,
\eqref{repaftrprojection} implies that there are three families,
transforming as
\begin{equation}
  (N,\bar{N},1)+(\bar{N},1,N)+(1,N,\bar{N})\label{reps}
\end{equation}
under the gauge group $SU(N)^3$. As demonstrated in
\eqref{projectedsuper}, the superpotential will remain the same,
including only the surviving fields. Finally, the differentiation of
the resulting unification groups occurs because of the different
ways the gauge group $SU(3N)$ is spontaneously broken. The minimal
models that are anomaly free are $SU(4)\times SU(2)\times SU(2)$,
$SU(4)^3$ and $SU(3)^3$\footnote{Similar approaches have been
examined in the framework of YM matrix models \cite{grosse-lizzi},
but they lacked phenomenological viability.}.

\subsection{An $SU(3)_c\times SU(3)_L\times SU(3)_R$ model}

Let us focus on the last option of the above breaking of $SU(3N)$,
which is the trinification group $SU(3)_c\times SU(3)_L\times
SU(3)_R$ \cite{glashow, rizov} (see also
\cite{ma-mondragon-zoupanos, lazarides-panagiotakopoulos, MaZoup,
babu-he-pakvasa, leontaris-rizos} and for a string theory approach
see \cite{kim}). At first, the integer $N$ has to be written as the
following decomposition
\begin{equation}
  N=n+3\,.
\end{equation}
Then, for $SU(N)$ the embedding
\begin{equation}
  SU(N)\supset SU(n)\times SU(3)\times U(1)\,,\label{decomp}
\end{equation}
is considered, from which it is found that the embedding for the
full gauge group $SU(N)^3$ is
\begin{equation}
  SU(N)^3\supset SU(n)\times SU(3)\times SU(n)\times SU(3)\times
  SU(n)\times SU(3)\times U(1)^3\,.\label{decomposition}
\end{equation}
The three $U(1)$ factors\footnote{These may be anomalous gaining
mass by the Green-Schwarz mechanism and therefore they decouple at
the low energy sector of the theory \cite{lawrence-nekrason-vafa}.}
are ignored and the representations \eqref{reps} are decomposed
according to the above decomposition, \eqref{decomposition}, (after
reordering the factors) as
\begin{align}
  &SU(n)\times SU(n)\times SU(n)\times SU(3)\times SU(3)\times
  SU(3)\,,\nonumber\\
 &(n,\bar{n},1;1,1,1)+(1,n,\bar{n};1,1,1)+(\bar{n},1,n;1,1,1)+(1,1,1;3,\bar{3},1)\nonumber\\
 &+(1,1,1;1,3,\bar{3})+(1,1,1;\bar{3},1,3)+(n,1,1;1,\bar{3},1)+(1,n,1;1,1,\bar{3})\nonumber\\
 &+(1,1,n;\bar{3},1,1)+(\bar{n},1,1;1,1,3)+(1,\bar{n},1;3,1,1)+(1,1,\bar{n};1,3,1)\,.
\end{align}
So, taking into account the decomposition \eqref{decomp}, the gauge
group is broken to $SU(3)^3$. Under the gauge group $SU(3)^3$, the
surviving fields transform according to
\begin{align}
  &SU(3)\times SU(3)\times SU(3)\,,\\
  &\left((3,\bar{3},1)+(\bar{3},1,3)+(1,3,\bar{3})\right)
\end{align}
representations, which correspond to the desired chiral
representations of the trinification group. Under the gauge group
$SU(3)_c\times SU(3)_L\times SU(3)_R$, the quarks and leptons of the
first family transform as
\begin{align}
  q=\left(
      \begin{array}{ccc}
        d & u & h \\
        d & u & h \\
        d & u & h \\
      \end{array}
    \right)\,\sim\,&(3,\bar{3},1)\,,\,\quad q^c=\left(
                                             \begin{array}{ccc}
                                               d^c & d^c & d^c \\
                                               u^c & u^c & u^c \\
                                               h^c & h^c & h^c \\
                                             \end{array}
                                           \right)\,\sim\,(\bar{3},1,3)\,,\,\nonumber\\
                                           \lambda&=\left(
                                                     \begin{array}{ccc}
                                                       N & E^c & \text{v} \\
                                                       E & N^c & e \\
                                                       \text{v}^c & e^c & S \\
                                                     \end{array}
                                                   \right)\sim(1,3,\bar{3})\,,
\end{align}
respectively. In a similar way, the matrices for the fermions of the
other two families are obtained.

It is crucial to note that this theory can be upgraded to a two-loop
finite theory (for reviews see \cite{inspire2, inspire4,
heinemeyer18, MaZoup}) and moreover it is able to make testable
predictions \cite{MaZoup}. Furthermore, fuzzy orbifolds can be used
in order to break spontaneously the unification gauge group down to
MSSM and afterwards to the $SU(3)_c\times U(1)_{em}$.

Summing up, we should emphasize the general picture of the
theoretical model. At very high-scale regime, we have an unbroken
renormalizable gauge theory. After the spontaneous symmetry
breaking, the resulting gauge theory is an $SU(3)^3$ GUT,
accompanied by a finite tower of massive Kaluza-Klein modes.
Finally, the trinification $SU(3)^3$ GUT breaks down to MSSM in the
low scale regime. Therefore, we conclude that fuzzy extra dimensions
can be used in constructing chiral, renormalizable and
phenomenologically viable field-theoretical models.

A natural extension of \thee above ideas \andd methods have been
reported in ref \cite{chatzi-stein-zoup} (see also
\cite{chatzi-stein-zoup2}), realized in \thee context of Matrix
Models (MM). At a fundamental level, \thee MMs introduced \by
Banks-Fischler-Shenker-Susskind (BFSS) \andd
Ishibashi-Kawai-Kitazawa-Tsuchiya (IKKT), are supposed to provide a
non-perturbative definition of M-theory \andd type IIB string theory
respectively \cite{ishibasi-kawai, banks}. On \thee other hand, MMs
are also useful laboratories for \thee study of structures which
could be relevant from a low-energy point of view. Indeed, they
generate a plethora of interesting solutions, corresponding to
strings, D-branes \andd their interactions \cite{ishibasi-kawai,
chepelev}, \as well \as to non-commutative/fuzzy spaces, such \as
fuzzy tori \andd spheres \cite{iso}. Such backgrounds naturally give
rise to non-abelian gauge theories. Therefore, it appears natural to
pose \thee question whether it is possible to construct
phenomenologically interesting particle physics models in this
framework \as well. In addition, an orbifold MM was proposed \by
Aoki-Iso-Suyama (AIS) in \cite{aoki} \as a particular projection of
\thee IKKT model, \andd it is directly related to \thee construction
described above in which fuzzy extra dimensions arise with
trinification gauge theory \cite{fuzzy}. By $\mathbb{Z}_3$ -
orbifolding, \thee original symmetry of \thee IKKT matrix model with
matrix size $3N\times 3N$ is generally reduced from $SO(9,1)\times
U(3N)$ to $SO(3,1) \times U(N)^3$. This model is chiral \andd has
$D=4$, $\mathcal{N}=1$ supersymmetry of Yang-Mills type \as well \as
an inhomogeneous supersymmetry specific to matrix models. The
$\mathbb{Z}_3$ - invariant fermion fields transform \as
bi-fundamental representations under \thee unbroken gauge symmetry
exactly \as in \thee constructions described above. In \thee future
we plan to extend further \thee studies initiated in refs
\cite{chatzi-stein-zoup, chatzi-stein-zoup2} in \thee context of
orbifolded IKKT models.\\

\emph{Acknowledgement}

\noindent This research is implemented under the Research Funding
Program ARISTEIA, Higher Order Calculations and Tools for High
Energy Colliders, HOCTools and the ARISTEIA II, Investigation of
certain higher derivative term field theories and gravity models
(co-financed by the European Union (European Social Fund ESF) and
Greek national funds through the Operational Program Education and
Lifelong Learning of the National Strategic Reference Framework
(NSRF)), as well as the European Unions ITN programme HIGGSTOOLS.

\end{document}